\pdfoutput=1
\documentclass[12pt,a4paper]{article}

\usepackage{ifthen} 
\usepackage{rotating}
\newboolean{pdflatex}
\setboolean{pdflatex}{true} 
\newboolean{articletitles}
\setboolean{articletitles}{true} 

\newboolean{uprightparticles}
\setboolean{uprightparticles}{false} 

\def\paperauthors{LHCb collaboration} 
\def\paperasciititle{Study of muon-tagged Ds1(2460) and Ds1(2536) decays to the Ds pi pi final state} 
\def\papertitle{Study of muon-tagged $D_{s1}(2460)^+$ and $D_{s1}(2536)^+$ decays to the $D_s^{+}\pi^+\pi^-$ final state} 
\def\paperkeywords{{High Energy Physics}, {LHCb}} 
\def\papercopyright{\the\year\ CERN for the benefit of the LHCb collaboration} 
\def\paperlicence{CC BY 4.0 licence}
\def\paperlicenceurl{https://creativecommons.org/licenses/by/4.0/}

\newif\ifEnableSectionTOCLinks
\EnableSectionTOCLinksfalse 

\usepackage[top=1in, bottom=1.25in, left=1in, right=1in]{geometry}

\usepackage{microtype}
\usepackage{lineno}  
\usepackage{xspace} 
\usepackage{caption} 

\usepackage{graphicx}  
\usepackage{color}
\usepackage{colortbl}
\graphicspath{{./figs/}} 

\usepackage{amsmath} 
\usepackage{amssymb}
\usepackage{amsfonts}
\usepackage{upgreek} 

\newcommand*\patchAmsMathEnvironmentForLineno[1]{%
\expandafter\let\csname old#1\expandafter\endcsname\csname #1\endcsname
\expandafter\let\csname oldend#1\expandafter\endcsname\csname
end#1\endcsname
 \renewenvironment{#1}%
   {\linenomath\csname old#1\endcsname}%
   {\csname oldend#1\endcsname\endlinenomath}%
}
\newcommand*\patchBothAmsMathEnvironmentsForLineno[1]{%
  \patchAmsMathEnvironmentForLineno{#1}%
  \patchAmsMathEnvironmentForLineno{#1*}%
}
\AtBeginDocument{%
\patchBothAmsMathEnvironmentsForLineno{equation}%
\patchBothAmsMathEnvironmentsForLineno{align}%
\patchBothAmsMathEnvironmentsForLineno{flalign}%
\patchBothAmsMathEnvironmentsForLineno{alignat}%
\patchBothAmsMathEnvironmentsForLineno{gather}%
\patchBothAmsMathEnvironmentsForLineno{multline}%
\patchBothAmsMathEnvironmentsForLineno{eqnarray}%
}

\usepackage[pdftex,
            pdfauthor={\paperauthors},
            pdftitle={\paperasciititle},
            pdfkeywords={\paperkeywords},
            ]{hyperref}

\usepackage{hyperxmp}

\usepackage[colorinlistoftodos,textsize=scriptsize]{todonotes}

\usepackage[bottom,flushmargin,hang,multiple]{footmisc}

\usepackage[all]{hypcap} 

\usepackage{xspace} 
\usepackage{upgreek}

\def\lhcb   {\mbox{LHCb}\xspace}

\def\MagUp {\mbox{\em Mag\kern -0.05em Up}\xspace}

\ifthenelse{\boolean{uprightparticles}}%
{

 \def\Pmu         {\ensuremath{\upmu}\xspace}                 
 \def\Pnu         {\ensuremath{\upnu}\xspace}                 
                  
 \def\Ppi         {\ensuremath{\uppi}\xspace}

 \def\Ppsi        {\ensuremath{\uppsi}\xspace}

 \def\PDelta      {\ensuremath{\Delta}\xspace}                 
 \def\PXi         {\ensuremath{\Xi}\xspace}                 
 \def\PLambda     {\ensuremath{\Lambda}\xspace}                 
 \def\PSigma      {\ensuremath{\Sigma}\xspace}                 
 \def\POmega      {\ensuremath{\Omega}\xspace}                 
 \def\PUpsilon    {\ensuremath{\Upsilon}\xspace}
 \let\oldPi\Pi
 \def\PPi         {\ensuremath{\oldPi}\xspace}

 \def\PB      {\ensuremath{\mathrm{B}}\xspace}                 
 \def\PD      {\ensuremath{\mathrm{D}}\xspace}                 
 \def\PJ      {\ensuremath{\mathrm{J}}\xspace}                 
 \def\PK      {\ensuremath{\mathrm{K}}\xspace}                 
 \def\Pb      {\ensuremath{\mathrm{b}}\xspace}                 
 \def\Pc      {\ensuremath{\mathrm{c}}\xspace}

 \def\Pp      {\ensuremath{\mathrm{p}}\xspace}                 

 \def\Ps      {\ensuremath{\mathrm{s}}\xspace}

 \def\thebaroffset{0.0em}
}
{

 \def\Pmu         {\ensuremath{\mu}\xspace}                 
 \def\Pnu         {\ensuremath{\nu}\xspace}                 
                  
 \def\Ppi         {\ensuremath{\pi}\xspace}

 \def\Ppsi        {\ensuremath{\psi}\xspace}                 
                  
 \mathchardef\PDelta="7101
 \mathchardef\PXi="7104
 \mathchardef\PLambda="7103
 \mathchardef\PSigma="7106
 \mathchardef\POmega="710A
 \mathchardef\PUpsilon="7107
 \mathchardef\PPi="7105
 \def\PB      {\ensuremath{B}\xspace}                 
 \def\PD      {\ensuremath{D}\xspace}                 
 \def\PJ      {\ensuremath{J}\xspace}                 
 \def\PK      {\ensuremath{K}\xspace}                 
 \def\Pb      {\ensuremath{b}\xspace}                 
 \def\Pc      {\ensuremath{c}\xspace}

 \def\Pp      {\ensuremath{p}\xspace}                 

 \def\Ps      {\ensuremath{s}\xspace}

 \def\thebaroffset{0.18em}
}
\newcommand{\offsetoverline}[2][\thebaroffset]{\kern #1\overline{\kern -#1 #2}}%

\makeatletter
\ifcase \@ptsize \relax
  \newcommand{\miniscule}{\@setfontsize\miniscule{4}{5}}
\or
  \newcommand{\miniscule}{\@setfontsize\miniscule{5}{6}}
\or
  \newcommand{\miniscule}{\@setfontsize\miniscule{5}{6}}
\fi
\makeatother

\DeclareRobustCommand{\optbar}[1]{\shortstack{{\miniscule (\rule[.5ex]{1.25em}{.18mm})}
  \\ [-.7ex] $#1$}}

\def\mun        {{\ensuremath{\Pmu^-}}\xspace} 

\def\neub       {{\ensuremath{\overline{\Pnu}}}\xspace}

\def\neumb      {{\ensuremath{\neub_\mu}}\xspace}

\def\squark    {{\ensuremath{\Ps}}\xspace}

\def\cquark    {{\ensuremath{\Pc}}\xspace}

\def\bquark    {{\ensuremath{\Pb}}\xspace}

\def\pion   {{\ensuremath{\Ppi}}\xspace}
\def\piz    {{\ensuremath{\pion^0}}\xspace}
\def\pip    {{\ensuremath{\pion^+}}\xspace}
\def\pim    {{\ensuremath{\pion^-}}\xspace}
\def\pipm   {{\ensuremath{\pion^\pm}}\xspace}

\def\kaon    {{\ensuremath{\PK}}\xspace}

\def\KorKbar {\kern \thebaroffset\optbar{\kern -\thebaroffset \PK}{}\xspace}

\def\Kp      {{\ensuremath{\kaon^+}}\xspace}
\def\Km      {{\ensuremath{\kaon^-}}\xspace}

\def\Kstar   {{\ensuremath{\kaon^*}}\xspace}

\def\Dbar    {{\ensuremath{\offsetoverline{\PD}}}\xspace}
\def\D       {{\ensuremath{\PD}}\xspace}
\def\Db      {{\ensuremath{\Dbar}}\xspace}
\def\DorDbar {\kern \thebaroffset\optbar{\kern -\thebaroffset \PD}\xspace}

\def\Dzb     {{\ensuremath{\Dbar{}^0}}\xspace}
\def\Dp      {{\ensuremath{\D^+}}\xspace}
\def\Dm      {{\ensuremath{\D^-}}\xspace}

\def\DpDm    {\ensuremath{\Dp {\kern -0.16em \Dm}}\xspace}
\def\Dstar   {{\ensuremath{\D^*}}\xspace}
\def\Dstarb  {{\ensuremath{\Dbar{}^*}}\xspace}

\def\Ds      {{\ensuremath{\D^+_\squark}}\xspace}
\def\Dsp     {{\ensuremath{\D^+_\squark}}\xspace}

\def\B       {{\ensuremath{\PB}}\xspace}
\def\Bbar    {{\ensuremath{\offsetoverline{\PB}}}\xspace}

\def\BorBbar {\kern \thebaroffset\optbar{\kern -\thebaroffset \PB}\xspace}

\def\Bd      {{\ensuremath{\B^0}}\xspace}

\def\BdorBdbar {\kern \thebaroffset\optbar{\kern -\thebaroffset \Bd}\xspace}
\def\Bu      {{\ensuremath{\B^+}}\xspace}

\def\Bp      {{\ensuremath{\Bu}}\xspace}

\def\Bs      {{\ensuremath{\B^0_\squark}}\xspace}
\def\Bsb     {{\ensuremath{\Bbar{}^0_\squark}}\xspace}
\def\BsorBsbar {\kern \thebaroffset\optbar{\kern -\thebaroffset \Bs}\xspace}

\def\jpsi     {{\ensuremath{{\PJ\mskip -3mu/\mskip -2mu\Ppsi}}}\xspace}

\def\Y#1S{\ensuremath{\PUpsilon{(#1S)}}\xspace}

\def\proton      {{\ensuremath{\Pp}}\xspace}

\def\Lz          {{\ensuremath{\PLambda}}\xspace}

\def\LorLbar     {\kern \thebaroffset\optbar{\kern -\thebaroffset \PLambda}\xspace}

\def\Lc          {{\ensuremath{\Lz^+_\cquark}}\xspace}

\def\Lb           {{\ensuremath{\Lz^0_\bquark}}\xspace}

\def\BF         {{\ensuremath{\mathcal{B}}}\xspace}
\def\BR         {\BF}

\def\to                 {\ensuremath{\rightarrow}\xspace}

\def\AT#1     {\ensuremath{A_{\mathrm{T}}^{#1}}\xspace}           

\def\C#1      {\ensuremath{\mathcal{C}_{#1}}\xspace}                       
\def\Cp#1     {\ensuremath{\mathcal{C}_{#1}^{'}}\xspace}                    
\def\Ceff#1   {\ensuremath{\mathcal{C}_{#1}^{\mathrm{(eff)}}}\xspace}        
\def\Cpeff#1  {\ensuremath{\mathcal{C}_{#1}^{'\mathrm{(eff)}}}\xspace}       
\def\Ope#1    {\ensuremath{\mathcal{O}_{#1}}\xspace}                       
\def\Opep#1   {\ensuremath{\mathcal{O}_{#1}^{'}}\xspace}                    

\newcommand{\nospaceunit}[1]{\ensuremath{\text{#1}}}       
\newcommand{\aunit}[1]{\ensuremath{\text{\,#1}}}       

\newcommand{\tev}{\aunit{Te\kern -0.1em V}\xspace}
\newcommand{\gev}{\aunit{Ge\kern -0.1em V}\xspace}
\newcommand{\mev}{\aunit{Me\kern -0.1em V}\xspace}
\newcommand{\kev}{\aunit{ke\kern -0.1em V}\xspace}
\newcommand{\ev}{\aunit{e\kern -0.1em V}\xspace}
 
\newcommand{\mevc}{\ensuremath{\aunit{Me\kern -0.1em V\!/}c}\xspace}
\newcommand{\gevc}{\ensuremath{\aunit{Ge\kern -0.1em V\!/}c}\xspace}
\newcommand{\mevcc}{\ensuremath{\aunit{Me\kern -0.1em V\!/}c^2}\xspace}
\newcommand{\gevcc}{\ensuremath{\aunit{Ge\kern -0.1em V\!/}c^2}\xspace}

\def\mum  {\ensuremath{\,\upmu\nospaceunit{m}}\xspace}

\def\fb   {\ensuremath{\aunit{fb}}\xspace}
\def\invfb   {\ensuremath{\fb^{-1}}\xspace}

\newcommand{\stat}{\aunit{(stat)}\xspace}
\newcommand{\syst}{\aunit{(syst)}\xspace}

\newcommand{\chisq}{\ensuremath{\chi^2}\xspace}

\newcommand{\chisqip}{\ensuremath{\chi^2_{\text{IP}}}\xspace}

\def\gsim{{~\raise.15em\hbox{$>$}\kern-.85em
          \lower.35em\hbox{$\sim$}~}\xspace}
\def\lsim{{~\raise.15em\hbox{$<$}\kern-.85em
          \lower.35em\hbox{$\sim$}~}\xspace}

\newcommand{\Real}{\ensuremath{\mathcal{R}e}\xspace}
\newcommand{\Imag}{\ensuremath{\mathcal{I}m}\xspace}

\def\sPlot{\mbox{\em sPlot}\xspace}

\def\pt         {\ensuremath{p_{\mathrm{T}}}\xspace}

\def\ptot       {\ensuremath{p}\xspace}

\def\evtgen     {\mbox{\textsc{EvtGen}}\xspace}

\def\geant      {\mbox{\textsc{Geant4}}\xspace}

\def\photos     {\mbox{\textsc{Photos}}\xspace}

\def\pythia     {\mbox{\textsc{Pythia}}\xspace}

\def\tensorflow {\mbox{\textsc{TensorFlow}}\xspace}

\def\tell1  {TELL1\xspace}
\def\ukl1   {UKL1\xspace}

\newcommand{\eg}{\mbox{\itshape e.g.}\xspace}

\newcommand{\lhcborcid}[1]{\href{https://orcid.org/#1}{\hspace*{0.1em}\raisebox{-0.45ex}{\includegraphics[width=1em]{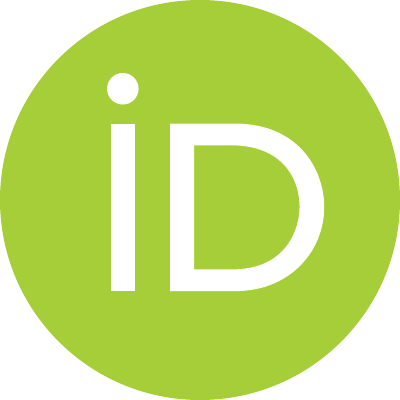}}}}

\hypersetup{
  colorlinks   = true, 
  urlcolor     = blue, 
  linkcolor    = blue, 
  citecolor    = red   
}

\ifEnableSectionTOCLinks
    \usepackage[explicit]{titlesec} 
    
    \let\oldcontentsline\contentsline
    \renewcommand

    \titleformat{\section}{\normalfont\Large\bf}{\hyperlink{tocsection.\thesection}{{\thesection} \parbox[t]{\dimexpr\textwidth-1pc}{#1}}}{1pc}{}

    \titleformat{\subsection}{\normalfont\bf}{\hyperlink{tocsubsection.\thesubsection}{{\thesubsection} \parbox[t]{\dimexpr\textwidth-1pc}{#1}}}{1pc}{}

    \titleformat{name=\section,numberless}[display]{}{}{0pt}{\normalfont\Huge\bfseries #1}
\fi

\usepackage{cite} 
\usepackage{mciteplus}

\usepackage{longtable} 
\usepackage{booktabs}

\usepackage{tikz}
\usetikzlibrary{decorations.markings}

\def\Dsone {\ensuremath{\D_{s1}^+}\xspace}
\def\Dsonep {\ensuremath{\D_{s1}(2460)^{+}}\xspace}
\def\Dsoneprp {\ensuremath{\D_{s1}(2536)^{+}}\xspace}
\def\Dsdecay {\mbox{$\Dsonep\to\Dsp\pip\pim$}\xspace}
\def\Dsdecaypr {\mbox{$\Dsoneprp\to\Dsp\pip\pim$}\xspace}
\def\Ddecaypr {\mbox{$\Dsoneprp\to\Dp\Kp\pim$}\xspace}
\def\Bsdecaypr {\mbox{$\Bsb\to\Dsoneprp\mun\neumb$}\xspace}
\def\Bdecay {\mbox{$\B\to\Dsonep\Db^{(*)}$}\xspace}
\def\Dspp {\mbox{$\Dsp\pip\pim$}\xspace}

\begin{document}

\renewcommand{\thefootnote}{\fnsymbol{footnote}}
\setcounter{footnote}{1}


\begin{titlepage}
\pagenumbering{roman}

\vspace*{-1.5cm}
\centerline{\large EUROPEAN ORGANIZATION FOR NUCLEAR RESEARCH (CERN)}
\vspace*{1.5cm}
\noindent
\begin{tabular*}{\linewidth}{lc@{\extracolsep{\fill}}r@{\extracolsep{0pt}}}
\ifthenelse{\boolean{pdflatex}}
{\vspace*{-1.5cm}\mbox{\!\!\!\includegraphics[width=.14\textwidth]{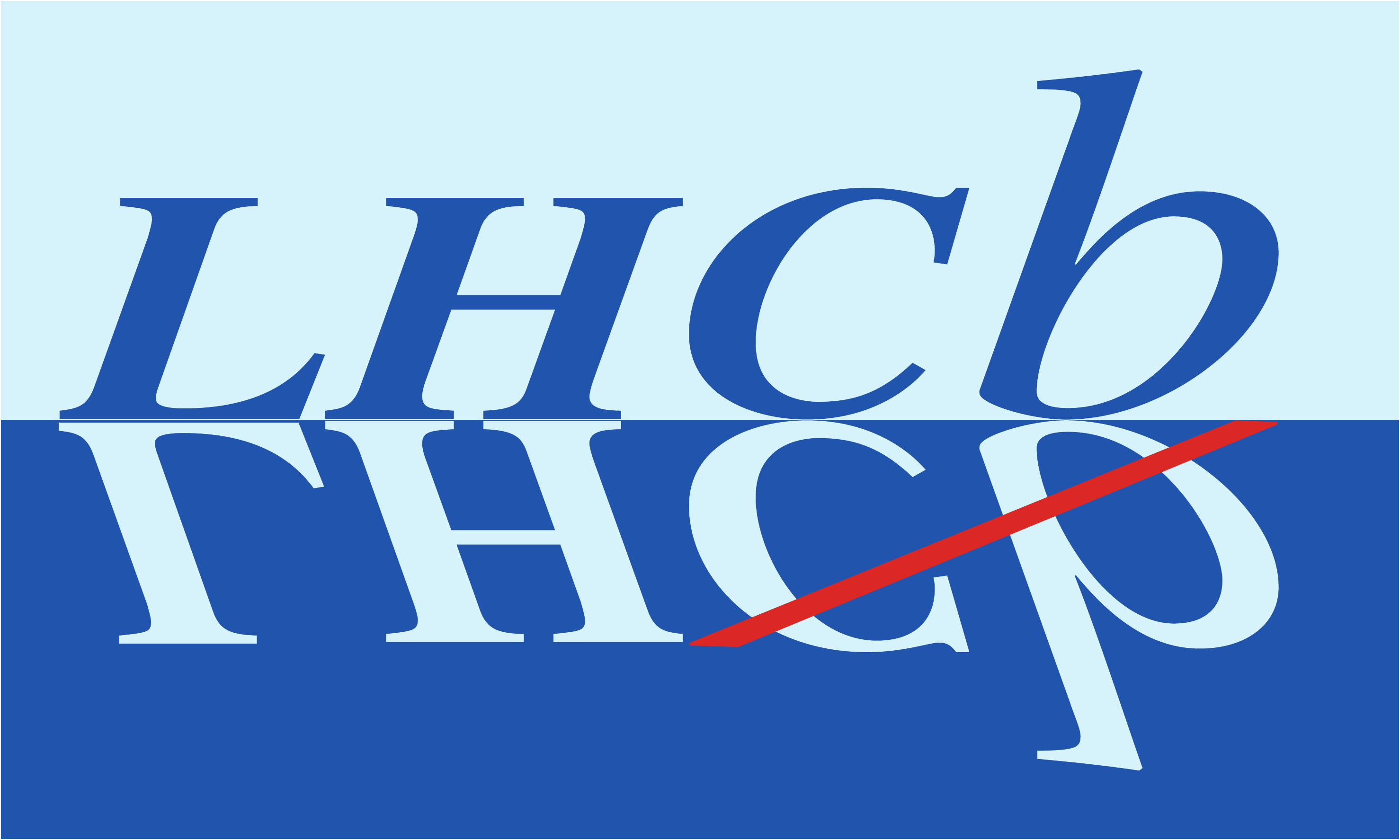}} & &}%
{\vspace*{-1.2cm}\mbox{\!\!\!\includegraphics[width=.12\textwidth]{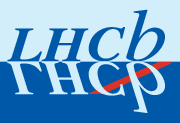}} & &}%
\\
 & & CERN-EP-2026-208 \\  
 & & LHCb-PAPER-2026-019 \\  
 & & 11 August 2026 \\ 
 & & \\
\end{tabular*}

\vspace*{3.0cm}

{\normalfont\bfseries\boldmath\huge
\begin{center}
  \papertitle 
\end{center}
}

\vspace*{1.0cm}

\begin{center}
\paperauthors\footnote{Authors are listed at the end of this paper.}
\end{center}

\vspace{\fill}

\begin{abstract}
  \noindent
  Decays of the pseudovector $D_{s1}(2460)^+$ and $D_{s1}(2536)^+$ mesons to the three-body $D_{s}^+\pi^+\pi^-$ final state are studied. The data sample is based on decays of beauty hadrons into $D_{s1}^+$ states accompanied by a muon from the $\bquark$-hadron decay chain collected by the LHCb detector during 2016--2018, corresponding to an integrated luminosity of 5.4 fb${}^{-1}$. The \mbox{$D_{s1}(2536)^+\to D_s^+\pi^+\pi^-$} branching fraction is measured for the first time, with the $D_{s1}(2536)^+\to D^+K^+\pi^-$ decay used as a reference. A simultaneous amplitude analysis of the $D_{s1}(2460)^+$ and $D_{s1}(2536)^+\to D_s^+\pi^+\pi^-$ decays is performed. The Dalitz-plot distributions of the two decays are found to be significantly different, suggesting differences in the internal structure of the two states, with evidence of exotic contributions to the $D_{s}^+\pi^{\pm}$ channel with the pole below the $DK$ threshold. Measurements of the masses of the $D_{s1}(2460)^+$ and $D_{s1}(2536)^+$ states are performed, and an upper limit on the $D_{s1}(2460)^+$ width is set.
\end{abstract}

\vspace*{1.5cm}

\begin{center}
  Submitted to
  JHEP
\end{center}

\vspace{\fill}

{\footnotesize 
\centerline{\copyright~\papercopyright. \href{\paperlicenceurl}{\paperlicence}.}}
\vspace*{2mm}

\end{titlepage}


\newpage
\setcounter{page}{2}
\mbox{~}
%
%
%
%


\renewcommand{\thefootnote}{\arabic{footnote}}
\setcounter{footnote}{0}


\cleardoublepage


\pagestyle{plain} 
\setcounter{page}{1}
\pagenumbering{arabic}


\section{Introduction}
\label{sec:Introduction}

Studies of \B-meson decays over the past few years have led to the discovery of a series of new four-quark states. For example, the LHCb collaboration reported the observation of the $T^*_{c\bar{s}0}(2900)^{++}$ and $T^*_{c\bar{s}0}(2900)^{0}$ mesons~\cite{LHCb-PAPER-2022-026, LHCb-PAPER-2022-027} with masses around 2.9\gev, decaying into $\Dsp\pipm$ final states.\footnote{Natural units with $\hbar=c=1$ are used, and the inclusion of charge-conjugate states is implied throughout this paper.} The quantum numbers of these states agree with $I(J^P) = 1(0^+)$. The proximity of their masses to the $\D^* K^*$ threshold\footnote{The omitted charged or neutral superscript in the notation means that both charged and neutral states are implied and isospin symmetry is assumed.} suggests that these particles might be $\D^* K^*$ bound states. If so, it is natural to expect an isospin-1 scalar molecular state near the $\D K$ threshold, decaying into $\Ds\pipm$. The width of such a state should reach 100--200\mev, since the decay into $\Ds\pi$ preserves isospin, in contrast to the $\Dsp\piz$ decay of the narrow isospin-0 $D_{s0}^*(2317)^+$ resonance. 
Therefore, it will be challenging to observe this state in $\B\to\Dbar\Ds\pi$ decays in the presence of other intermediate states~\cite{LHCb-PAPER-2022-027}. An alternative way to search for an isospin-1 state near the $\D K$ threshold could be in decays of the pseudovector $\Dsonep$ meson to $\Dsp \pip \pim$. 
A further motivation to investigate \Dsdecay decays is the study of the nature of the $\Dsonep$ state itself~\cite{Tang:2023yls}, where a molecular $\Dstar K$ component can affect the Dalitz-plot distribution of its decays. 

An alternative interpretation of the $T^*_{c\bar{s}0}(2900)$ structures, along with the $T^*_{cs0}(2870)$ state observed with the LHCb experiment earlier in the $D^+K^-$ channel~\cite{LHCb-PAPER-2020-024, LHCb-PAPER-2020-025}, is provided by the open-charm tetraquark model of Ref.~\cite{Maiani:2024quj}, where the observed states form part of an SU(3) flavour multiplet. In this picture, the signals around 2.9\gev are interpreted as radially excited isospin-1 tetraquarks, which are too heavy to be accommodated as ground states. The same framework relates these structures to the $D_{s0}^*(2317)^+$ meson, which is assigned to an isosinglet with negative strangeness, and predicts a corresponding ground-state isotriplet with masses around 2.3\gev that could decay to the $\Dsp\pi$ final state. 

The analysis of \Dsdecay decays produced in \Bdecay transitions has recently been performed by LHCb~\cite{LHCb-PAPER-2024-033}, where the structure of the Dalitz plot suggested the presence of $\Dsp\pipm$ states with a pole near to the $\D K$ threshold. 
An interpretation of these results involving only $\pip\pim$ states is also possible, although unlikely: \eg, it requires a significant contribution of the tensor $f_2(1270)$ state with the pole far outside of the kinematic region. 

Studying the decays of the other pseudovector state, $\Dsoneprp$, to the same final state could help clarify the exotic or conventional interpretation of the structures seen in the \Dsonep decay. Indeed, the study presented in Ref.~\cite{LHCb-PAPER-2024-033} led to predictions of the $\Dsoneprp\to\Dsp\pip\pim$  decay amplitude~\cite{Dias:2025izv, Yang:2025dcg}, where the \Dsoneprp state is assumed to be either a conventional $c\bar{s}$ meson or a $\D^* K$ molecule. In particular, the predicted $\pip\pim$ invariant-mass distributions differ significantly between these two interpretations, which provides further motivation to study the $\Dsoneprp\to\Dsp\pip\pim$ decays. 

This paper is dedicated to the study of the $\Dsp\pip\pim$ final state from decays of $\Dsonep$ and $\Dsoneprp$ mesons produced in $\B$-meson decays tagged by an accompanying muon from the $\B$-meson decay chain. Unlike the $B$ decays producing two charm hadrons (double-charm $B$ decays) used in Ref.~\cite{LHCb-PAPER-2024-033}, such muon-tagged decays provide abundant samples of both $\Dsonep$ and $\Dsoneprp$ states. The production mechanism of the orbitally excited $\Ds$ states in $\B$-meson decays depends on the total angular momentum $\vec{j}$ of the light $\squark$ quark, and is thus different for \Dsonep and \Dsoneprp. In semileptonic $b\to c\ell\nu$ transitions, the production of $j=1/2$ states (scalar $\D_{s0}^*(2317)^+$ and pseudovector $\Dsonep$) is suppressed with respect to the $j=3/2$ states (pseudovector $\Dsoneprp$ and tensor $\D_{s2}^*(2573)^+$) by the smallness of the Isgur-Wise function $\tau_{1/2}$ compared to $\tau_{3/2}$~\cite{Morenas:1997nk, Becirevic:2012te}. As a result, the muon-tagged \Dsonep state is predominantly produced in \Bdecay decays with subsequent semileptonic $\Db\to \mun \neumb X$ decays, while the $\Dsoneprp$ arises mostly from semileptonic \Bsdecaypr transitions. The optimal selections for the two states are thus different. 

The dynamics of the \Dsdecay and \Dsdecaypr decays is studied by analysing their Dalitz-plot distributions. In addition, the previously unmeasured branching fraction of \Dsdecaypr relative to the known \Ddecaypr decay is determined, and an updated measurement of the \Dsonep and \Dsoneprp masses and the width of the $\Dsonep$ state is presented. 
\section{Detector and simulation}
\label{sec:detector}

The \lhcb detector~\cite{LHCb-DP-2008-001,LHCb-DP-2014-002} is a single-arm forward spectrometer covering the \mbox{pseudorapidity} range $2<\eta <5$, designed for the study of particles containing \bquark or \cquark quarks. The detector used to collect the data analysed in this paper includes a high-precision tracking system consisting of a silicon-strip vertex detector surrounding the proton-proton~($pp$) interaction region~\cite{LHCb-DP-2014-001}, a large-area silicon-strip detector located upstream of a dipole magnet with a bending power of about $4{\mathrm{\,T\,m}}$, and three stations of silicon-strip detectors and straw drift tubes~\cite{LHCb-DP-2017-001} placed downstream of the magnet. The tracking system provides a measurement of the momentum, \ptot, of charged particles with a relative uncertainty that varies from 0.5\% at low momentum to 1.0\% at 200\gev. The minimum distance of a track to a primary $pp$ collision vertex (PV), the impact parameter (IP), is measured with a resolution of $(15+29/\pt)\mum$, where \pt is the component of the momentum transverse to the beam, in\,\gev. Different types of charged hadrons are distinguished using information from two ring-imaging Cherenkov (RICH) detectors~\cite{LHCb-DP-2012-003}.  Photons, electrons and hadrons are identified by a calorimeter system consisting of scintillating-pad and preshower detectors, an electromagnetic and a hadronic calorimeter. Muons are identified by a system composed of alternating layers of iron and multiwire proportional chambers~\cite{LHCb-DP-2012-002}.

The online event selection is performed by a trigger~\cite{LHCb-DP-2012-004}, which consists of a hardware stage, based on information from the calorimeter and muon systems, followed by a software stage, which applies a full event reconstruction. At the hardware trigger stage, events are required to have a muon with high \pt or a hadron, photon or electron with high transverse energy in the calorimeters. The software trigger requires a two-, three- or four-track secondary vertex with a significant displacement from any PV. At least one charged particle must have a high $\pt$ and be inconsistent with originating from any PV. A multivariate algorithm~\cite{LHCb-PROC-2015-018} is used for the identification of secondary vertices consistent with the decay of a \bquark hadron.

Simulation is required to model the effects of the detector acceptance and the imposed selection requirements. In the simulation, $pp$ collisions are generated using \pythia~\cite{Sjostrand:2007gs} with a specific \lhcb configuration~\cite{LHCb-PROC-2010-056}. Decays of unstable particles are described by \evtgen~\cite{Lange:2001uf}, in which final-state radiation is generated using \photos~\cite{davidson2015photos}. The interaction of the generated particles with the detector, and its response, are implemented using the \geant toolkit~\cite{Allison:2006ve, *Agostinelli:2002hh} as described in Ref.~\cite{LHCb-PROC-2011-006}. The simulated samples are corrected to account for known data-simulation differences in track reconstruction, particle identification (PID) and trigger efficiencies.

\section{Signal selection}
\label{sec:selection}

The analysis is performed using data collected by the LHCb detector in 2016--2018 in \textit{pp} collisions at a centre-of-mass energy of 13\tev, and corresponding to an integrated luminosity of 5.4\invfb. The $\Dsonep$ and $\Dsoneprp$ candidates reconstructed in the \Dspp final state are used in the amplitude analysis. Additionally, \Ddecaypr decays are selected as a control channel and as a normalisation sample for the \Dsdecaypr branching-fraction measurement. 

The $\Dsp$ candidates are reconstructed in the \mbox{$\Kp\Km\pip$} final state, and the $\Dp$ candidates in the \mbox{$\Km\pip\pip$} final state. Pion and kaon tracks used to form the $\D_{(s)}^+$ candidates are selected based on loose requirements on track-fit quality, $p$, and \pt of the tracks. In addition, they must be within the RICH detector acceptance and positively identified as pions or kaons, respectively. To suppress background from random combinations of tracks originating at the $\proton\proton$ interaction vertices, all tracks are required to be displaced from any PV in the event. The displacement is characterised by the quantity $\chisqip$, defined as the difference in the vertex-fit \chisq of a given PV reconstructed with and without the track under consideration. To suppress backgrounds from misidentified \mbox{$\Lc \to \proton \Km \pip$} decays, the positively charged tracks forming the $\D_{(s)}^+$ candidates must not be identified as protons.

For the $\Dsone\mun$ pairs selection, where $\Dsone$ denotes either of the \Dsonep and \Dsoneprp states, the muon candidates are required to have $p \in [
3,100]\gev$, and to be identified as muons. The cosine of the angle between the combined momentum of the $\D_{s1}^+ \mun$ system and the line connecting the PV to the $\D_{s1}^+ \mun$ decay vertex is required to exceed 0.999. The $\Dsone\mun$ and $\D_{(s)}^+$ vertex fits are required to have good quality. The invariant mass of the $\Dsp$ ($\Dp$) candidates is required to be within $\pm 12\mev$ ($\pm 15\mev$) of the known mass of the corresponding particle~\cite{PDG2024}, and the reconstructed $\Dsone$ mass must be below $2700\mev$. The invariant mass of the full $\Dsone\mun$ system must lie between $2500$ and $5200\mev$. 

In the offline selection, trigger signals are associated with reconstructed particles. Selection requirements are applied on the trigger selection itself and on whether the decision was due to the signal candidate, other particles produced in the $pp$ collision, or a combination of both.
Each $\Dsone\mun$ candidate must belong to at least one of three categories with respect to the hardware-trigger decision, which cover most of the possible trigger scenarios: the hadron trigger decision is due to the tracks of the $\D_{(s)}^+$ candidate, the muon trigger decision is due to the muon candidate, or the event is triggered by other particles in the $\proton\proton$ collision, unrelated to the $\Dsone\mun$ candidate. 

Additional background suppression is achieved by reconstructing the missing mass squared \mbox{$m^2_{\rm miss} \equiv (p_{\B}-p_{\D_{s1}}-p_{\mu})^2$}, where the four-momenta $p_{\D_{s1}}$ of the $\D_{s1}^+$ candidate and $p_{\mu}$ of the muon are reconstructed directly. In contrast, the four-momentum of the $\B$ meson $p_{\B}$ is calculated approximately by assuming that the boost of the $\B$ candidate is equal to that of the visible $\Dsone\mun$ combination~\cite{LHCb-PAPER-2015-025}. The $m^2_{\rm miss}$ variable peaks near zero for semileptonic candidates, therefore the selection optimised for the semileptonic $\Dsoneprp$ signal requires $-1 < m^2_{\rm miss} < 2\gev^2$, while that for the double-charm $\Dsonep$ signal uses a looser requirement of $-0.5 < m^2_{\rm miss} < 4.5\gev^2$ to account for additional missing particles in the decay.

To further enhance the purity of the \mbox{$\D_{s1}^+\to \Dsp\pip\pim$} signal, a multivariate selection based on a boosted decision tree (BDT)~\cite{Breiman} implemented with the  {\mbox{\textsc{Scikit-learn}}\xspace} package~\cite{Scikit-learn-paper} is also applied. 
Two classifiers are trained separately. One is optimised to be sensitive to \Dsdecay decays produced in double-charm decays, and is referred to as DD selection. The other is optimised for \Dsdecaypr decays produced in semileptonic $\B$ meson decays and is referred to as SL selection. 
As input observables, the BDT uses kinematic and topological quantities, including the quality of the kinematic fit of the decay chain, flight-distance significances, \chisqip values, \pt, and PID responses for the final-state tracks. In both cases, simulated events are used as signal proxy, and candidates from the sidebands of the $m(\Dsp\pip\pim)$ distribution in data as background proxy. The training samples are equally split into the training and test samples to control overtraining. 
The effect of overtraining due to the use of sideband samples is evaluated and is found to be negligible. 
The working points are chosen to retain about $60\%$ of the signal, corresponding to a background retention rate of $\sim 10\%$ ($\sim 5\%$) for the DD (SL) selection, which provides a good compromise between statistical precision and robustness of the subsequent amplitude analysis. No BDT selection is applied for the $\Dsoneprp\to\Dp\Kp\pim$ channel.

A small fraction of events contains more than one reconstructed signal candidate after the final selection. In the mass fits used to determine signal yields, such multiple candidates are retained, since they contribute to the smooth combinatorial background. In the \Dspp Dalitz-plot analysis, where most of the duplicate candidates differ by a single pion exchange, they could distort the density of events across phase space. Only a single arbitrarily chosen candidate is thus kept in such events.

\section{Invariant-mass distributions and Dalitz plots}
\label{sec:massfit}

\begin{figure}
    \centering
    \includegraphics[width=0.5\linewidth]{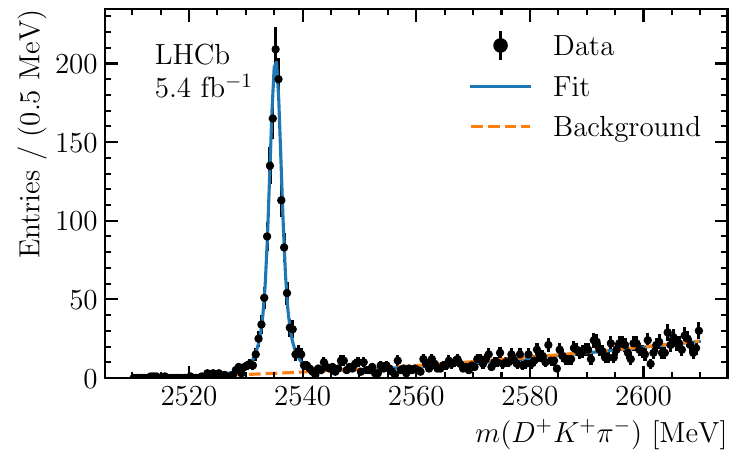}
    \caption{Invariant-mass distribution of $\Dp\Kp\pim$ candidates from the $\Dp\Kp\pim\mun$ sample, with the fit result detailed in the text overlaid.}
    \label{fig:dkpi_massfit}
\end{figure}

The invariant-mass distribution of the muon-tagged $\Dp\Kp\pim$ combinations after the full selection is shown in Fig.~\ref{fig:dkpi_massfit}, with the fit superimposed. The distribution is modelled by the sum of three components: the \Ddecaypr signal, a potential \mbox{$\D^{\ast}_{s2}(2573)^{+}\to\Dp\Kp\pim$} contribution, and a smooth combinatorial component. The \Ddecaypr signal is parametrised by a function obtained from simulation, which is a convolution of a relativistic Breit--Wigner (RBW) function with the \Dsoneprp width fixed to the known value~\cite{PDG2024} and a resolution function represented by a sum of two Gaussian distributions sharing the same central value. The $\D^{\ast}_{s2}(2573)^{+}\to\Dp\Kp\pim$ component is parametrised by a Breit--Wigner distribution with the mass and width of the $\D^{\ast}_{s2}(2573)^{+}$ state fixed to the known values~\cite{PDG2024}. The combinatorial component is parametrised with a second-order Bernstein polynomial. The parameters allowed to vary in the fit are the $\Dsoneprp$ and $\D^{\ast}_{s2}(2573)^{+}$ yields, the width of the $\Dsoneprp$ state, its mass difference with respect to the value used in simulation and the Bernstein polynomial parameters. 

The fitted \Ddecaypr yield is $1340\pm 40$, and the signal peak position corresponds to a \Dsoneprp mass of $2535.18\pm 0.04$\mev. The yield of the \mbox{$\D^{\ast}_{s2}(2573)^{+}\to\Dp\Kp\pim$} component in the fit is constrained to be positive and is found to be consistent with zero, with an upper limit of $140$ events at 90\% confidence level. 

\begin{figure}[b]
    \centering
    \includegraphics[width=0.49\linewidth]{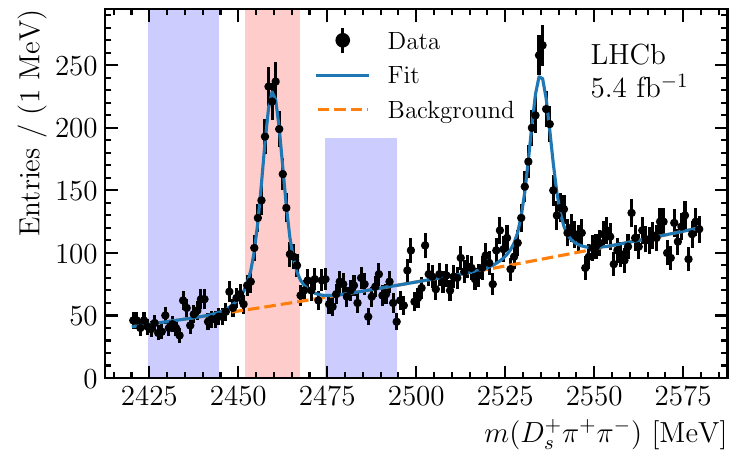}
    \put(-30,40){(a)}
    \includegraphics[width=0.49\linewidth]{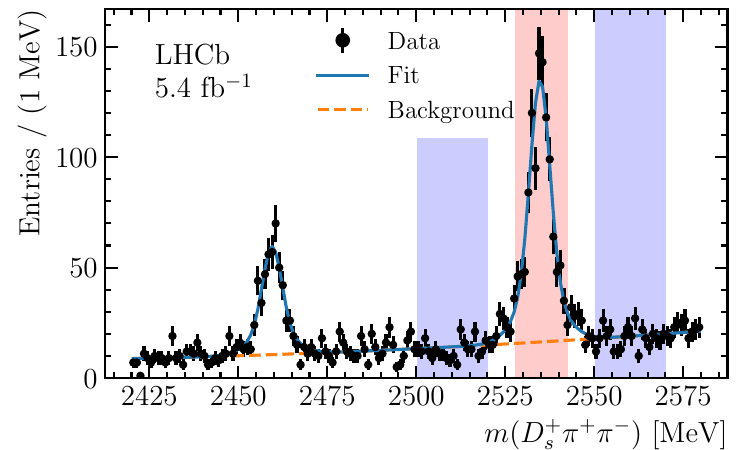}
    \put(-182,40){(b)}
    \caption{Invariant-mass distribution of $\Dsp\pip\pim$ candidates from the $\Dsp\pip\pim\mun$ sample after the (a) DD and (b) SL selections, with their fit results overlaid. Shaded regions illustrate signal and background-dominated regions defined in the text. }
    \label{fig:dspipi_massfit}
\end{figure}

The distributions of the \Dspp invariant mass with muon tags after the DD and SL selections are shown in Fig.~\ref{fig:dspipi_massfit}. Each distribution is parametrised by the sum of three components: two peaks corresponding to the \Dsonep and \Dsoneprp decays to \Dspp, and a smooth background due to random combinations of particles. Each signal peak is represented by the convolution of an RBW distribution and a resolution function parametrised as the sum of two Gaussian distributions with the same mean, with the widths and relative fractions of the two Gaussian functions fixed from simulation. 
Background components are represented by second-order Bernstein polynomials with floating parameters. 
A Gaussian constraint on the $\Dsoneprp$ Breit--Wigner width is applied in the fit, with the central value and its uncertainty taken from the world-average values~\cite{PDG2024}. The common scale factor for the widths of the resolution functions is allowed to vary in the fit to account for the possible mismodelling of the resolution in simulation. The other parameters free in the fit are the yields of the two signal peaks, their mass shifts with respect to the known values, and the $\Dsonep$ Breit--Wigner width. 
Shaded areas in Fig.~\ref{fig:dspipi_massfit} indicate the signal and the background-dominated (sideband) regions used in the Dalitz-plot analyses. The signal regions are defined as $\pm 7.7\mev$ ($\pm 7.4\mev$) intervals around the fitted \Dsonep (\Dsoneprp) mass peak, corresponding to approximately $\pm 2.5$ times the invariant-mass resolution. The sideband regions are defined as \mbox{$15<|m(\Dsp\pip\pim)-m_0|<35\mev$}, where $m_0$ is the fitted peak position of the corresponding $\D_{s1}^+$ state.

\begin{table}[]
    \centering
    \caption{Fit results for the $\Dsp\pip\pim$ invariant-mass distributions from the DD and SL selections, as well as the signal ($N_{\rm sig}$) and background ($N_{\rm bkg}$) yields in the signal invariant-mass regions for each of the \Dsone states. The uncertainties are statistical only.}
    \label{tab:dspipi_massfit}
    \begin{tabular}{lcc}
\toprule
               Parameter & DD selection & SL selection \\
\midrule
  $D_{s1}(2460)^+$ yield & $1360 \pm 70$ & $376 \pm 25$ \\
  $D_{s1}(2536)^+$ yield & $1180 \pm 60$ & $980 \pm 40$ \\
        $m_{2460}$ [MeV] & $2459.51 \pm 0.14\phantom{0}$ & $2459.39 \pm 0.23\phantom{00}$ \\
        $m_{2536}$ [MeV] & $2534.71 \pm 0.18\phantom{0}$ & $2534.68 \pm 0.15\phantom{00}$ \\
\midrule
  $N_{\rm sig}(\Dsonep)$ & $1240 \pm 60$ & $357 \pm 23$ \\
  $N_{\rm bkg}(\Dsonep)$ & $836 \pm 8$ & $155.2 \pm 3.4\phantom{0}$ \\
$N_{\rm sig}(\Dsoneprp)$ & $1125 \pm 60$ & $921 \pm 40$ \\
$N_{\rm bkg}(\Dsoneprp)$ & $1315 \pm 12$ & $226 \pm 5\phantom{0}$ \\
\bottomrule
\end{tabular}

\end{table}

The fit results for the $\Dsp\pip\pim$ invariant-mass distributions for both SL and DD selections are summarised in Table~\ref{tab:dspipi_massfit}. The signal and background yields in the signal regions for each of the $\D_{s1}^+$ states are also given. The mass of the \Dsoneprp state from the $\Dsp\pip\pim$ fit differs from that obtained from the $\Dp\Kp\pim$ fit by about three standard deviations, however the two measurements are compatible when the systematic uncertainty due to potential interference with nonresonant components, discussed in Sec.~\ref{sec:branching}, is taken into account. The fitted value of the $\Dsonep$ width from the DD sample is $0.62\pm 0.54\mev$. The upper limits on the width are found to be 1.26\mev and 1.38\mev at 90\% and 95\% confidence levels, respectively. These values include the dominant systematic uncertainties due to the resolution modelling and the $\Dsoneprp$ width. 

\begin{figure}
    \centering
    \includegraphics[width=0.4\linewidth]{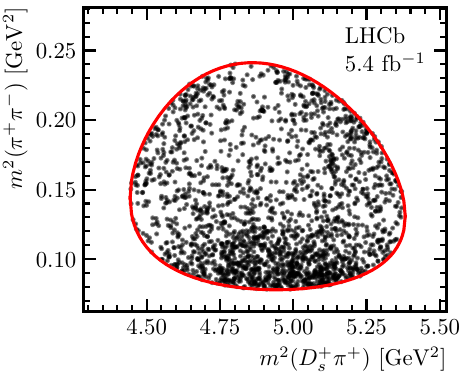}
    \put(-140,122){(a)}
    \includegraphics[width=0.4\linewidth]{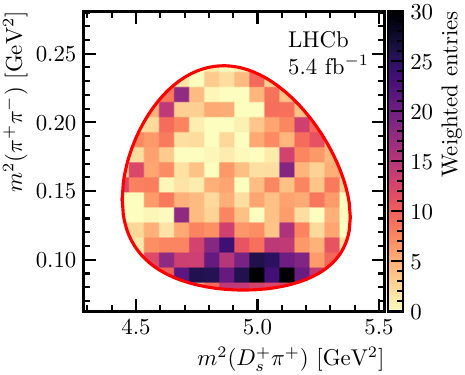}
    \put(-140,122){(b)}
    
    \includegraphics[width=0.4\linewidth]{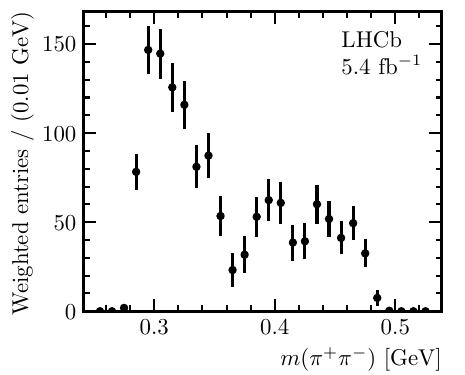}
    \put(-142,38){(c)}
    \includegraphics[width=0.4\linewidth]{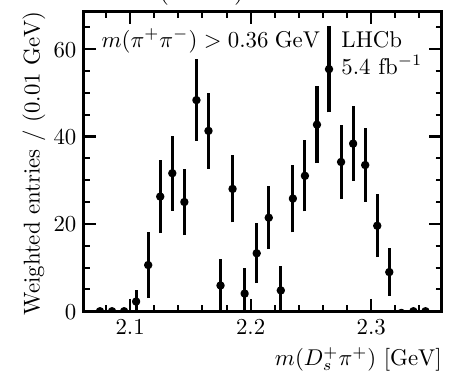}
    \put(-142,38){(d)}
  \caption{Dalitz-plot distributions for the \Dsdecay sample: (a) raw distribution; (b) background-subtracted and efficiency-corrected binned distributions, with its (c) $m(\pip\pim)$ and (d) $m(\Dsp\pip)$ projections for candidates with $m(\pip\pim)>0.36\gev$.}
  \label{fig:dlz1}
\end{figure}

\begin{figure}
    \centering
    \includegraphics[width=0.4\linewidth]{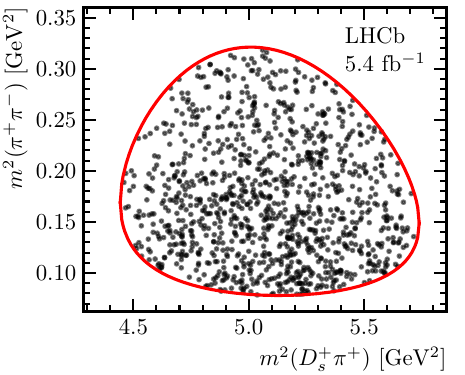}
    \put(-140,122){(a)}
    \includegraphics[width=0.4\linewidth]{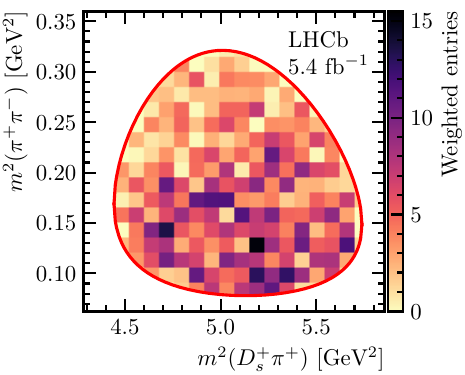}
    \put(-140,122){(b)}
    
    \includegraphics[width=0.4\linewidth]{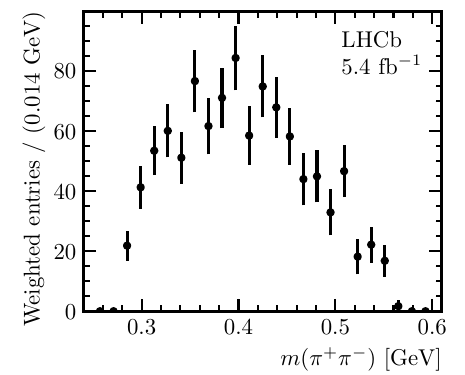}
    \put(-144,35){(c)}
    \includegraphics[width=0.4\linewidth]{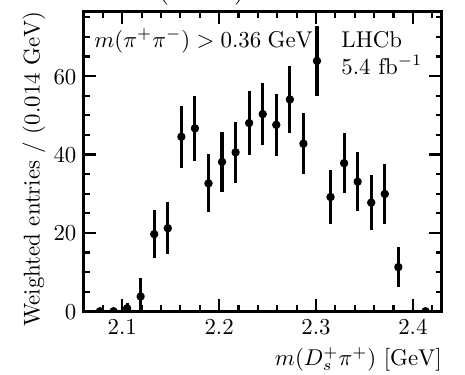}
    \put(-144,35){(d)}
      \caption{Dalitz-plot distributions for the \Dsdecaypr sample: (a) raw distribution; (b) background-subtracted and efficiency-corrected binned distributions, with its (c) $m(\pip\pim)$ and (d) $m(\Dsp\pip)$ projections for candidates with $m(\pip\pim)>0.36\gev$.}
    \label{fig:dlz2}
\end{figure}

The Dalitz plots of the three-body \mbox{$\D_{s1}^+\to \Dsp\pip\pim$} decays studied in this
analysis are shown in Figs.~\ref{fig:dlz1}
and \ref{fig:dlz2}. The Dalitz-plot variables are obtained from a
kinematic fit~\cite{Hulsbergen:2005pu} in which the masses of the initial $\Dsone$
and of the final-state $\Dsp$ mesons are fixed to their known
values~\cite{PDG2024}. The one-dimensional projections and two-dimensional binned
distributions are corrected for nonuniform efficiency and the 
background is subtracted in these plots using the \sPlot technique~\cite{Pivk:2004ty} with the 
invariant-mass fits described above.

The \Dsdecay Dalitz plot, shown in Fig.~\ref{fig:dlz1}, exhibits 
pronounced structures consistent with those seen in Ref.~\cite{LHCb-PAPER-2024-033}. 
The $m(\pip\pim)$ distribution has a clear enhancement near threshold and a
minimum around $0.36\gev$. The $m(\Dsp\pip)$ distribution, which is
strongly anticorrelated with $m(\Dsp\pim)$ owing to the almost
degenerate phase space, shows a single peak for events with $m(\pip\pim)<0.36\gev$ but a
distinct double-peak structure for $m(\pip\pim)>0.36\gev$. Such
behaviour cannot be described by a uniform phase-space model or by a
pure $S$-wave $\pip\pim$ contribution, such as a single $f_0(500)$
resonance.

In contrast, the \Dsdecaypr Dalitz plot in Fig.~\ref{fig:dlz2} is much more uniform. In particular, the double-peak structure observed at high $m(\pip\pim)$ in the \Dsdecay sample is not seen here, which indicates a different dynamical composition of the two amplitudes despite the same quantum numbers of the initial states and identical final states.

\section{Amplitude formalism}
\label{sec:formalism}

The $\Dsone$ mesons studied in this analysis are produced in $\bquark$-hadron decays accompanied by a muon that is either coming from the semimuonic decay of the other charm hadron in double-charm decays, such as \mbox{$\Bp\to\Dsone\Dzb$}, or is the result of a semimuonic decay of the $\bquark$ hadron itself, such as \mbox{$\Bs\to\Dsone\mun\neumb$}. Due to unreconstructed particles in these decay chains, one cannot access the complete kinematic information of the $\bquark$-hadron decay. The $\Dsone$ mesons are assumed to be unpolarised in this analysis, and the kinematics of their decays is therefore characterised by two-dimensional Dalitz plots without adding the polarisation degrees of freedom as was done in Ref.~\cite{LHCb-PAPER-2024-033}. 

The distribution of the three-body $\Dsone$ decay is described by the Mandelstam variables $s_i$, defined as $s_1 = m^2_{23} = m^2(\pip\pim)$, $s_2 = m^2_{13} = m^2(\Dsp\pim)$, and $s_3 = m^2_{12} = m^2(\Dsp\pip)$, of which only two are independent. The indices used to label the particles in the decay are: 1 for the $\Dsp$, 2 for the $\pip$, and 3 for the $\pim$. 
The $\Dsone\to\Dsp\pip\pim$ decay density over the Dalitz-plot phase space $\Omega=(s_1, s_3)$ is obtained by summing the amplitudes incoherently over the three polarisation states of the $\Dsone$ meson: 
\begin{equation}
  \frac{{\rm d}\Gamma}{{\rm d}\Omega} = \sum\limits_{\nu=-1,0,1}|\mathcal{O}_{\nu}(s_1, s_3)|^2, 
  \label{eq:decay_density}
\end{equation}
where the Dalitz-plot function $\mathcal{O}_{\nu}$ is parametrised following the Dalitz-plot decomposition formalism~\cite{JPAC:2019ufm}. The index $\nu$ denotes the component of the $\Dsone$ spin quantised along the $\pip\pim$ direction. 

The function $\mathcal{O}_{\nu}(s_1, s_3)$ is represented as a coherent sum of $K$ amplitude components $\mathcal{O}^{(k)}_{\nu}(s_1, s_3) = a_{k,\nu} A_{k,\nu}(s_1, s_3)$ ($1<k<K$), corresponding to intermediate states of different quantum numbers in different channels, where $a_{k,\nu}$ is the complex coupling and the amplitude $A_{k,\nu}$ describes the dynamical and angular dependence of the component $k$. The intermediate states considered here are scalar ($J^P=0^+$) and tensor ($J^P=2^+$) $\pip\pim$ amplitudes, as well as scalar ($T^*_{c\bar{s}0}$) and vector ($T^*_{c\bar{s}1}$) amplitudes in the $\Dsp\pip$ and $\Dsp\pim$ channels. The amplitude models used in the analysis and the corresponding lineshape parametrisations are summarised below.

\subsection{Helicity amplitudes}
\label{sec:helicity_amplitudes}

Following Ref.~\cite{JPAC:2019ufm}, the amplitude for the decay of a $\Dsone$ meson with quantum numbers $J^P=1^+$ to a $\Dsp$ meson ($J^P=0^-$) and an intermediate scalar $\pip\pim$ system ($J^P=0^+$, denoted $f_0$) can be written as
\begin{equation}
\mathcal{O}^{(f_0)}_{\nu}(s_1) = a_1\; 
\frac{p_1}{p^{(0)}_1}\; \mathcal{R}^{(f_0)}_1(s_1) \, \delta_{\nu,0},
\label{eq:ampl_sigma}
\end{equation}
where $\mathcal{R}_1^{(f_0)}(s_1)$ is the lineshape of the intermediate $f_0(500)$ state, $\delta_{\nu,0}$ is the Kronecker delta symbol, and $p_1/p^{(0)}_1$ is the $P$-wave threshold factor, included explicitly. The specific choice of the reference momentum $p_1^{(0)}$ only changes the numerical values of the couplings without affecting the physics conclusions. Here $p_1^{(0)}$ is the two-body $\D_{s1}\to\Dsp f_0$ breakup momentum evaluated at fixed reference masses ($0.4\gev$ for the resonance in the $\pip\pim$ channel). Since only relative couplings can be determined, the coupling $a_1$ corresponding to the $f_0(500)$ is fixed to unity. In the models where an additional scalar $f_0(980)$ is present, its coupling is a free parameter. 

For the intermediate tensor ($J^P=2^+$) $\pip\pim$ state, denoted $f_2$, the corresponding contribution is constructed analogously within the helicity formalism of Ref.~\cite{JPAC:2019ufm}, and added coherently to the scalar term. 

In the baseline model including intermediate states in the $\Dsp\pipm$ channels, the Dalitz-plot function $\mathcal{O}_{\nu}$ is written as a coherent sum of three decay chains with scalar ($J^P=0^+$) resonances in $\pip\pim$ given by Eq.~(\ref{eq:ampl_sigma}), plus the contribution of the $\Dsp\pip$ and $\Dsp\pim$ combinations, given by
\begin{equation}
        O^{(\Ds\pi)}_{\nu}(s_1, s_3) = a_2\; \frac{p_2}{p^{(0)}_2}\; \mathcal{R}_2(s_2) d^1_{\nu,0}(-\hat{\theta}_{1(2)}) + 
                                     a_3\; \frac{p_3}{p^{(0)}_3}\; \mathcal{R}_3(s_3) d^1_{\nu,0}(\hat{\theta}_{3(1)}). \\
    \label{eq:scalar_ampl}
\end{equation}
These terms contain Wigner rotations with angles $\hat{\theta}_{i(j)}$ that align the decay chains with the intermediate $\Dsp\pipm$ combinations to the quantisation axis with the resonance in the $\pip\pim$ channel. The explicit expressions for these angles as functions of the Dalitz plot variables can be found in Ref.~\cite{JPAC:2019ufm}.
Assuming isospin symmetry, the couplings and lineshapes are equal: $a_2 = a_3 \equiv a$ and $\mathcal{R}_2(s) = \mathcal{R}_3(s) \equiv \mathcal{R}(s)$. The reference breakup momenta $p_{2,3}^{(0)}$ are calculated at the $\Dsp\pipm$ invariant masses of $2.25\gev$. 

An alternative scenario assumes that the intermediate $\Dsp\pipm$ states are vectors (${J^P=1^-}$) produced in the combination of $S$- and $D$-waves and decaying via $P$-wave to $\Dsp\pipm$. The corresponding helicity amplitudes can again be constructed following Ref.~\cite{JPAC:2019ufm}; in this analysis they are used as auxiliary models, although they are not consistent with the double-charm analysis data~\cite{LHCb-PAPER-2024-033}.

\subsection{Resonance lineshapes}
\label{sec:lineshapes}

The functions $\mathcal{R}(s)$ encode the dynamics of the $\pip\pim$ and $\Dsp\pipm$ interactions. Several alternative parametrisations are considered and used in different combinations for systematic uncertainty studies, as summarised in Table~\ref{tab:lineshapes}.

\begin{table}[t]
  \centering
  \caption{Lineshape models used for $\pip\pim$ and $\Dsp\pipm$ amplitudes.}
  \label{tab:lineshapes}
  \begin{tabular}{lll}
    \toprule
    Channel & Resonances & Lineshape models \\
    \midrule
    $\pip\pim$ & $f_0(500)$, $f_0(980)$, $f_2(1270)$ & Breit--Wigner (baseline) \\
               & $f_0(500)$ & Quasi model-independent \\
    \midrule
    $\Dsp\pipm$ & $T^*_{c\bar{s}0}$, $T^*_{c\bar{s}1}$ & Breit--Wigner \\
                & $T^*_{c\bar{s}0}$                  & Scattering-length approximation (baseline) \\
                & $T^*_{c\bar{s}0}$                  & $\Dstar K$ and $\D\Kstar$ triangle diagrams \\
    \bottomrule
  \end{tabular}
\end{table}

The common approach to parametrise an intermediate resonance $R$ in Dalitz-plot analyses is to use the RBW function. In this analysis, it is used for both $\pip\pim$ and $\Dsp\pipm$ resonances, and takes the form
\begin{equation}
    \mathcal{R}_{\rm RBW}(m) = \frac{F_R(m,L_R)F_D(m,L_D)}{m_{0}^{2}-m^{2}-i m_{R}\Gamma(m)},
\end{equation}
with mass-dependent width
\begin{equation}
    \Gamma(m) = \Gamma_{R}\left(\frac{q(m)}{q^{(0)}}\right)^{2L_R+1}\frac{m_{R}}{m}F_{R}^{2}(m,L_R).
\end{equation}
Here, $m$ is the invariant mass of the resonance decay products, $m_{R}$ and $\Gamma_{R}$ the pole mass and width of the resonance, $L_D$ and $L_R$ the orbital angular momenta in the production and decay vertices, $q(m)$ the breakup momentum in the rest frame of the resonance with mass $m$, while $q^{(0)}$ is the $R$ breakup momentum at $m=m_R$. The Blatt--Weisskopf form factors $F_{R}(m,L_R)$ and $F_{D}(m,L_D)$ are given by
\begin{equation}
  F_{R,D}\left(m,L\right) = \begin{cases}
        1 & \mbox{for } L = 0, \\
        \sqrt{\dfrac{1+z^{2}(m)}{1+z_{0}^{2}}} & \mbox{for } L = 1,\\
        \sqrt{\dfrac{9+3z^{2}(m)+z^{4}(m)}{9+3z^{2}_{0}+z^{4}_{0}}} & \mbox{for } L = 2,\\
  \end{cases}
\end{equation}
with $z(m) = q(m)d$ and $z_{0}=q^{(0)}d$ for $F_R$, and $z(m) = p(m)d$ and $z_{0}=p^{(0)}d$ for $F_D$. Here $p(m)$ is the two-body momentum in $\Dsone\to R\Dsp$ or $\Dsone\to R\pipm$, and $p^{(0)}$ is defined as in Eq.~(\ref{eq:scalar_ampl}). The radial parameter is taken as $d=4\gev^{-1}$ for all resonances. 

Since the $\pip\pim$ invariant masses in the fits do not exceed 0.6\gev, a single Breit-Wigner function provides an adequate description of the $\pip\pim$ $S$--wave. An alternative, quasi model-independent (QMI) description is also considered, in which the complex amplitude is parametrised as a cubic spline with a small number of knots spanning the kinematic region. The knots are evenly spaced in $m(\pip\pim)$, with the first and last knots placed at the kinematic boundaries of the \Dsdecaypr decay. The real and imaginary parts of the amplitude at each knot are free parameters in the fit except for the central knot, where the amplitude is fixed to unity.

For the $\Dsp\pipm$ channels, a coupled-channel scattering-length approximation (SLA) is adopted to describe near-threshold molecular-like states 
(see, \eg, Ref.~\cite{Fernandez-Ramirez:2019koa} and Sec. 50.3.3 of Ref.~\cite{PDG2022}). The scattering amplitude matrix $\mathcal{M}$ for the $DK$ and $\Dsp\pi$ channels is written in $K$-matrix form as
\begin{equation}
    \mathcal{M} = \frac{\mathcal{K}}{1-i\mathcal{K}\rho},
\end{equation}
where $\rho$ is the diagonal matrix of dimensionless phase-space factors,
\begin{align}
    \rho &= \mathrm{diag}\bigl(\rho_{DK}(m),\, \rho_{D_s\pi}(m)\bigr), \\
    \rho_{AB}(m) &= \frac{1}{16\pi}\,\frac{\sqrt{[m^2-(m_{A}+m_{B})^2][m^2-(m_{A}-m_{B})^2]}}{m^2},
\end{align}
analytically continued below threshold using $\sqrt{x} = +i\sqrt{|x|}$. The constant $K$-matrix is
\begin{equation}
    \mathcal{K} = \begin{pmatrix} \gamma & \beta \\ \beta & \gamma_2 \end{pmatrix},
    \label{eq:k_matrix}
\end{equation}
where $\gamma$ is proportional to the elastic $DK$ scattering length, $\beta$ is responsible for the coupling between $DK$ and $\Dsp\pi$, and $\gamma_2$ encodes possible direct interaction in the $\Dsp\pi$ channel. In the baseline fit, $\gamma_2$ is set to zero, while variants with $\gamma_2$ free to float in the fit are used to assess systematic effects. The $\Dsp\pi$ lineshape entering the amplitude model is given by the $\mathcal{M}_{22}$ element,
\begin{equation}
    \mathcal{R}_{\rm SLA}(m) = \frac{\beta^{2} \,\rho_{DK}(m) + i\gamma_2 \bigl(i \gamma \,\rho_{DK}(m) - 1\bigr)}{\beta^{2} \,\rho_{DK}(m) \,\rho_{D_{s}\pi}(m) + \bigl(i \gamma \,\rho_{DK}(m) - 1\bigr) \bigl(i \gamma_2 \,\rho_{D_{s}\pi}(m) - 1\bigr)}.
\end{equation}

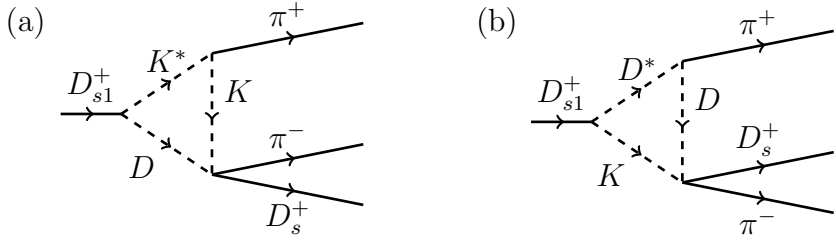
\begin{figure}
    \centering
    \begin{tikzpicture}[line width=1.0 pt, scale=0.4]
        \tikzset{onshell/.style={draw=black, postaction={decorate,
            decoration={markings, mark=at position .55 with {\arrow[black]{>}}}}}}
        \tikzset{offshell/.style={draw=black, dashed, postaction={decorate,
            decoration={markings, mark=at position .55 with {\arrow[black]{>}}}}}}

        \coordinate (Ds1) at (0,0);
        \coordinate (Ds) at (3,2);
        \coordinate (K) at (3,-2);
        
        \draw[onshell] (-2,0) -- (Ds1) node[midway, above] {$D_{s1}^+$};
        
        \draw[offshell] (Ds1) -- (Ds) node[midway, above] {$K^*$};
        \draw[offshell] (Ds) -- (K) node[midway, above right] {$K$};
        \draw[offshell] (Ds1) -- (K) node[midway, below left] {$D$};
        
        \draw[onshell] (Ds) -- (8,3) node[midway, above] {$\pi^+$};
        \draw[onshell] (K) -- (8,-1) node[midway, above] {$\pi^-$};
        \draw[onshell] (K) -- (8,-3) node[midway, below] {$\Dsp$};
    \end{tikzpicture}
    \put(-135, 80){(a)}
    \hspace{2cm}
    \begin{tikzpicture}[line width=1.0 pt, scale=0.4]
        \tikzset{onshell/.style={draw=black, postaction={decorate,
            decoration={markings, mark=at position .55 with {\arrow[black]{>}}}}}}
        \tikzset{offshell/.style={draw=black, dashed, postaction={decorate,
            decoration={markings, mark=at position .55 with {\arrow[black]{>}}}}}}

        \coordinate (Ds1) at (0,0);
        \coordinate (Ds) at (3,2);
        \coordinate (K) at (3,-2);
        
        \draw[onshell] (-2,0) -- (Ds1) node[midway, above] {$D_{s1}^+$};
        
        \draw[offshell] (Ds1) -- (Ds) node[midway, above] {$D^{*}$};
        \draw[offshell] (Ds) -- (K) node[midway, above right] {$D$};
        \draw[offshell] (Ds1) -- (K) node[midway, below left] {$K$};
        
        \draw[onshell] (Ds) -- (8,3) node[midway, above] {$\pi^+$};
        \draw[onshell] (K) -- (8,-1) node[midway, above] {$\Dsp$};
        \draw[onshell] (K) -- (8,-3) node[midway, below] {$\pi^-$};
    \end{tikzpicture}
    \put(-135, 80){(b)}
    \caption{Feynman diagrams for the \mbox{$\Dsone\to \Dsp\pip\pim$} decay mediated by a triangle loop with (a) intermediate $DK^{*}$ states (Tri$(\D \Kstar)$ amplitude) and (b) intermediate $\Dstar K$ states (Tri$(\Dstar K)$ amplitude). }
    \label{fig:triangle_diagram_sketch}
\end{figure}

As suggested \eg in Refs.~\cite{Dias:2025izv,Yang:2025dcg,Roca:2025lij}, the observed $\Dsp\pipm$ structures can be described by triangle loop diagrams with intermediate $\Dstar K$ (denoted here as Tri$(\Dstar K)$) or $\D\Kstar$ (denoted as Tri$(\D \Kstar)$) states (see Fig.~\ref{fig:triangle_diagram_sketch}). To test this hypothesis, the lineshapes from the LHCb analysis of $\Lb\to\jpsi\proton\Km$ decays~\cite{LHCb-PAPER-2019-014} are used for the $\Dsp\pipm$ channels. The principal difference of these models is that they do not feature a phase rotation in the \Dsonep kinematic region, while for the \Dsoneprp decay, the phase rotation is only present below (above) the $\D K$ threshold for $\Dstar K$ ($\D\Kstar$) loops. 

\subsection{Maximum-likelihood fit}

The amplitude analysis is performed with an unbinned
maximum-likelihood fit to the two-dimensional Dalitz-plot kinematics of
the $\Dsonep$ and $\Dsoneprp$ decays to \Dspp final state. Events in the
signal regions of the two $\Dsp\pip\pim$ invariant-mass peaks, defined in
Sec.~\ref{sec:massfit}, are used. The likelihood function $\mathcal{L}$ for a sample of $N$ candidates is given by
\begin{equation}
  -2\ln \mathcal{L} = -2\sum\limits_{i=0}^{N}\ln p_{\rm tot}(\Omega_i),
\end{equation}
where the total probability density including
signal, efficiency and background contributions is given by
\begin{equation}
  p_{\rm tot}(\Omega) = p_{\rm sig}(\Omega)\,\varepsilon(\Omega)
  \frac{1-f_{\rm bkg}}{\mathcal{N}_{\rm sig}} + p_{\rm bkg}(\Omega)\frac{f_{\rm bkg}}{\mathcal{N}_{\rm bkg}}.
  \label{eq:tot_density}
\end{equation}
Here $p_{\rm sig}(\Omega) = d\Gamma(\Omega)/d\Omega$ is the signal
Dalitz-plot density as a function of the phase-space coordinates
$\Omega$ defined in Eq.~(\ref{eq:decay_density}), $\varepsilon(\Omega)$
is the efficiency profile (see Sec.~\ref{sec:efficiency}), and $p_{\rm bkg}(\Omega)$ describes the
background distributions in the Dalitz-plot variables (see Sec.~\ref{sec:background}). The background
fraction $f_{\rm bkg}$ in the signal region is taken from the
$\Dsp\pip\pim$ invariant-mass fits of Sec.~\ref{sec:massfit}. The
normalisation factors are
\begin{equation}
  \mathcal{N}_{\rm sig} = \int p_{\rm sig}(\Omega)\, \varepsilon(\Omega)\,\mathrm{d}\Omega,\qquad
  \mathcal{N}_{\rm bkg} = \int p_{\rm bkg}(\Omega)\, \mathrm{d}\Omega,
\end{equation}
where the integrals extend over the full physical phase space. They are
evaluated numerically using samples of events distributed uniformly in the 
phase space. The absolute normalisation of the functions $p_{\rm sig}$, $p_{\rm bkg}$, and $\varepsilon$ is arbitrary in this definition. In simultaneous fits to the $\Dsonep$ and $\Dsoneprp$ samples, the combined negative log-likelihood
\begin{equation}
  -2\ln \mathcal{L}^{(\rm comb)} = -2\ln \mathcal{L}^{(2460)} -2\ln \mathcal{L}^{(2536)}
\end{equation}
is minimised.

The signal-decay density $p_{\rm sig}(\Omega)$ following the formalism described above 
is constructed using the \textsc{AmpliTF} package~\cite{amplitf}. 
The amplitude fit is implemented in \textsc{TFA2}~\cite{tfa2}, a fitting package based on \tensorflow framework~\cite{tensorflow2015-whitepaper}, interfaced with the \textsc{iminuit} minimisation library~\cite{iminuit}. 

The quality of the fits is characterised by a $\chisq$ computed over a uniform rectangular binning in $s_1$ and $s_3$ variables, with eight bins in the kinematically allowed regions of both variables. Both \Dsonep and \Dsoneprp Dalitz plots have 60 bins in the kinematically allowed region. The effective number of degrees of freedom for the $\chisq$ test is evaluated using pseudoexperiments and equals 54 to 56 for the models used in the study. 

\subsection{Fit fractions and interference}
\label{sec:fit_fractions}

In addition to the couplings $a_{k,\nu}$, fit fractions are reported for the amplitude analysis results as a more convention-independent measure of the relative importance of each component in the overall decay dynamics. The fit fraction $\mathcal{F}(k)$ of component $k$ is defined in this analysis as the
ratio of its intensity summed over the $\Dsone$ polarisation states to the 
total intensity over the Dalitz plot,
\begin{equation}
  \mathcal{F}(k) = \left(\sum_{\nu} \int |c_{k,\nu} A_{k,\nu}(\Omega)|^2\,\mathrm{d}\Omega\right)\left/
                  \left(\sum_{\nu} \int \biggl|\sum_{\ell=1}^{K} c_{\ell,\nu} A_{\ell,\nu}(\Omega)\biggr|^2\,\mathrm{d}\Omega\right).\right.
  \label{eq:fit_fraction}
\end{equation}
In the present analysis, the scalar $\pip\pim$ and $\Dsp\pipm$ 
amplitudes have very similar kinematic dependence with slow variations
of the dynamical lineshape functions $\mathcal{R}(s)$ over the limited
phase space of the \mbox{$\Dsone\to\Dsp\pip\pim$} decays. This leads to large interference
effects, resulting in sums of fit fractions that differ significantly from unity.

\section{Amplitude analysis}
\label{sec:aman}

The amplitude analysis aims at a coherent description of the decays
$\Dsonep$ and $\Dsoneprp$ to the $\Dsp\pip\pim$ final state using a
simultaneous fit to the two samples. If the two states are conventional
$c\bar s$ mesons, they are both admixtures of the states with total spin
of the two quarks $S=0$ and $S=1$. In the heavy-quark limit, the
\Dspp decay proceeds through the $S=0$ component for both
mesons, thus, since the masses of the two states are close, their decay
dynamics would be similar. The observed differences
between the \Dspp amplitudes of the two mesons can thus be
interpreted as evidence for a nontrivial structure of the initial
state, such as an admixture of a hadronic molecule.

A significant molecular $D^{*}K$ component in the $\D_{s1}^+$ wave function
would naturally generate a near-threshold $DK$ molecular configuration
after the decay of the $D^*$ meson. Through $DK\to\Ds\pi$ rescattering, this
would induce nontrivial behaviour in the $\Dsp\pipm$ structures of the
\Dspp Dalitz plot. This motivates, in addition to more
conventional descriptions, a dedicated study of a two-channel
rescattering model based on a $K$-matrix formalism.

\subsection{Efficiency}
\label{sec:efficiency}

The efficiency across the Dalitz plot is accounted for in the signal probability density in Eq.~\eqref{eq:tot_density} via the factor $\varepsilon(\Omega)$, and is determined from simulation. Separate samples are used for the different decay modes: double-charm $B$ decays for \Dsdecay and semileptonic \Bsdecaypr decays for \Dsdecaypr and \Ddecaypr. In each case, the detector acceptance, trigger, reconstruction and selection requirements are applied to the simulated events, and the efficiency is parametrised as a smooth function of the Dalitz-plot variables.

\begin{figure}
    \centering
    \includegraphics[width=0.47\linewidth]{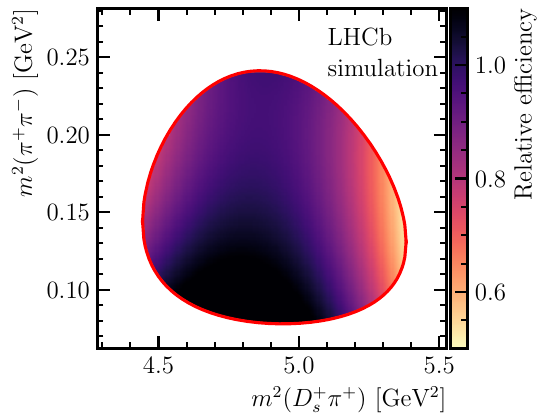}
    \put(-165,140){(a)}
    \includegraphics[width=0.47\linewidth]{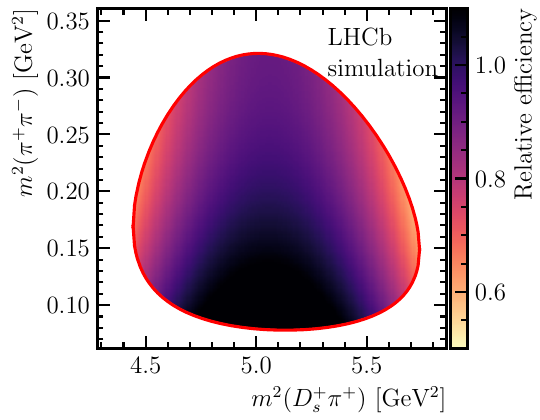}
    \put(-165,140){(b)}
    \caption{Relative efficiency profiles across the Dalitz plots for (a) \Dsdecay and (b) \Dsdecaypr decays.}
    \label{fig:efficiency}
\end{figure}

For the \mbox{$\Dsone\to\Dsp\pip\pim$} channels, $\varepsilon(\Omega)=\varepsilon(x,y)$ is described by a third-order polynomial in the two variables, $x$ and $y$, chosen here as $s_3 = m^2(\Dsp\pip)$ and $s_1 = m^2(\pip\pim)$, respectively, which are linearly transformed to the interval $(-1,1)$ to reduce correlations in the fit parameters. 
The overall normalisation of $\varepsilon(x,y)$ is arbitrary for the likelihood fit and is fixed by imposing $\varepsilon(0,0)=1$. The coefficients are obtained from unbinned maximum-likelihood fits to the simulated signal samples and are independent for the $\Dsonep$ and $\Dsoneprp$ decays. 
The resulting efficiency profiles are shown in Fig.~\ref{fig:efficiency}.
The chosen parametrisation is sufficiently flexible to reproduce the smooth variation in efficiency across the limited phase space of the decays, while avoiding overfitting to statistical fluctuations of the simulated samples.

\subsection{Background density}
\label{sec:background}

The Dalitz-plot background density $p_{\rm bkg}(\Omega)$ entering the likelihood of Eq.~(\ref{eq:decay_density}) is obtained from the same data sidebands in the three-body invariant-mass distributions as used for the multivariate selection (Sec.~\ref{sec:selection}). 
The background density is parametrised using fully connected neural networks with $L_2$ regularisation implemented in the \tensorflow framework, following the procedure outlined in Ref.~\cite{Mathad:2019rqj}. Separate networks are trained for each decay mode. The input layer takes the two Dalitz-plot coordinates, and two hidden layers with sigmoid activation functions (32 and 8 neurons, respectively) are used to model the smooth background variations. The networks are trained with the \textsc{Adam}\ optimiser~\cite{kingma2014adam} using sideband data as training samples, and the output is normalised to unity over the kinematic phase space using a large ensemble of uniformly generated phase-space points.

\begin{figure}
    \centering
    \includegraphics[width=0.47\linewidth]{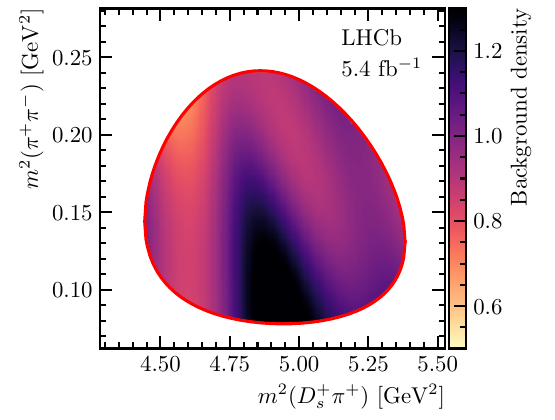}
    \put(-165,140){(a)}
    \includegraphics[width=0.47\linewidth]{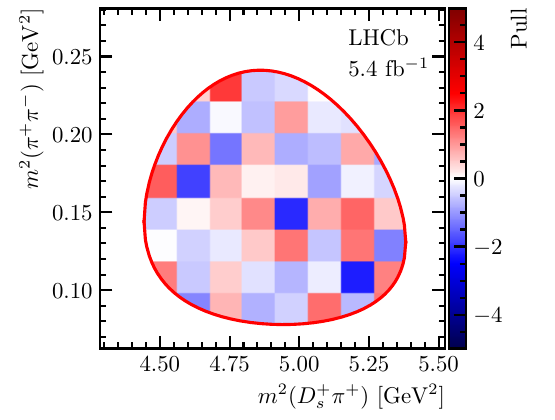}
    \put(-165,140){(b)}

    \includegraphics[width=0.47\linewidth]{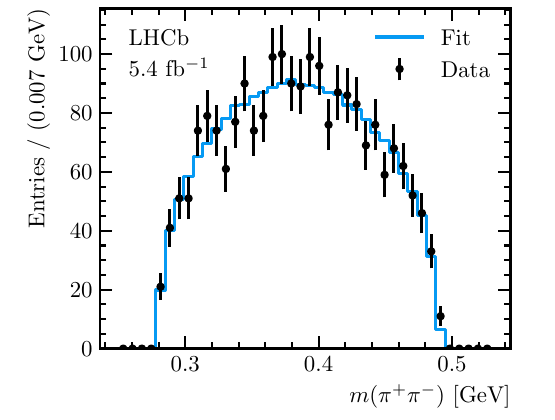}
    \put(-165,110){(c)}
    \includegraphics[width=0.47\linewidth]{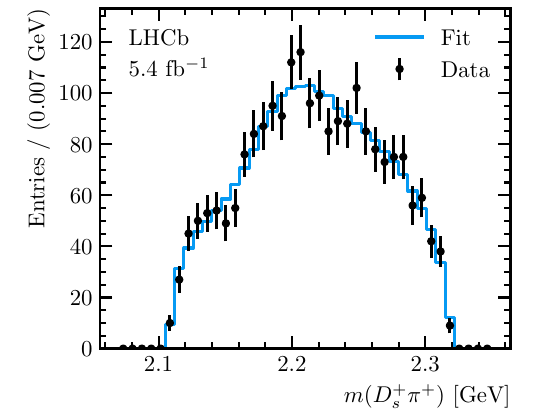}
    \put(-165,110){(d)}
    \caption{Dalitz-plot background distributions for the \Dsdecay sample: (a) two-dimensional density in the Dalitz plane, (b) its pull distribution, and (c,d) its one-dimensional projections. The points represent sideband data, and the histograms the corresponding neural-network model.}
    \label{fig:bck_dspipi1}
\end{figure}

\begin{figure}
    \centering
    \includegraphics[width=0.47\linewidth]{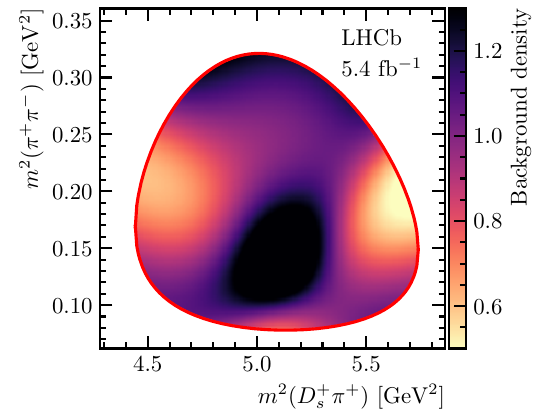}
    \put(-165,140){(a)}
    \includegraphics[width=0.47\linewidth]{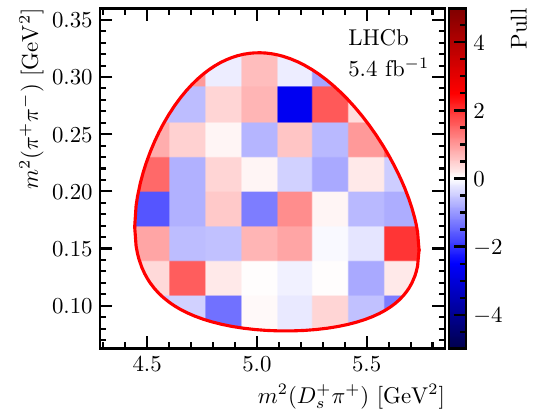}
    \put(-165,140){(b)}

    \includegraphics[width=0.47\linewidth]{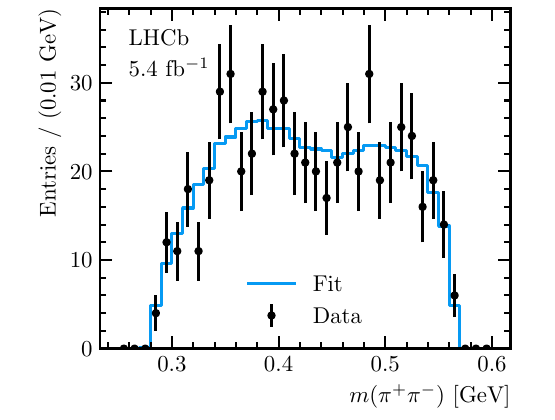}
    \put(-165,110){(c)}
    \includegraphics[width=0.47\linewidth]{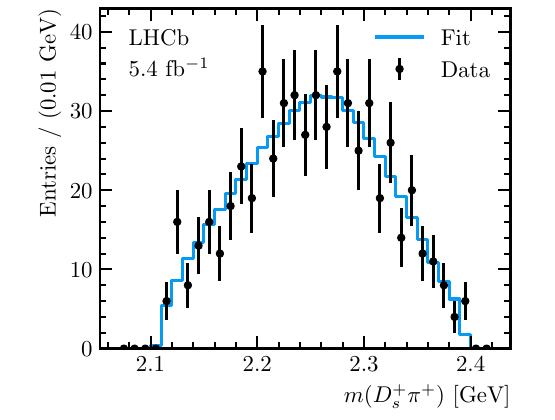}
    \put(-165,110){(d)}
    \caption{Dalitz-plot background distributions for the \Dsdecaypr sample: (a) two-dimensional density in the Dalitz plane, (b) its pull distribution, and (c,d) its one-dimensional projections. The points represent sideband data, and the histograms the corresponding neural-network model.}
    \label{fig:bck_dspipi2}
\end{figure}

The regularisation strength is chosen such that the background model reproduces the main structures observed in the Dalitz plots of the sideband regions without following statistical fluctuations. The resulting Dalitz-plot distributions and their one-dimensional projections are shown in Figs.~\ref{fig:bck_dspipi1} and \ref{fig:bck_dspipi2} for the \Dsdecay and \Dsdecaypr samples, respectively. 
Variations in the $L_2$ parameter around its nominal value are considered as part of the systematic uncertainty studies on the amplitude fit.

\subsection{Models with resonances in the \texorpdfstring{\boldmath{$\pip\pim$}}{pipi} channel}
\label{sec:app_resonances}

The features observed in the \Dsdecay Dalitz plot are
first tested against models that contain only $\pip\pim$ resonances.
Since a single scalar state in the $\pip\pim$ channel cannot reproduce
the structures seen in Fig.~\ref{fig:dlz1}, higher partial
waves are introduced. A low-$m(\pip\pim)$ contribution from the
$\rho(770)$ meson is isospin-violating in $D_{s1}^+\to\Dsp\rho(770)$ and 
therefore expected to be strongly suppressed. 
The broad isoscalar tensor state $f_2(1270)$, with
$\Gamma=186.6\pm2.3\mev$, may instead contribute through its low-mass
tail.

The corresponding amplitude model is defined in Sec.~\ref{sec:helicity_amplitudes}. 
Several variants of the $\pip\pim$ amplitude model are considered and are listed in Table~\ref{tab:model_summary}, which shows the negative log-likelihood 
values and the binned $\chi^2$ for the combined fits to the $\Dsonep$ and 
\Dsdecaypr samples. 
In the minimal Breit--Wigner model with a scalar $f_0(500)$ and a tensor
$f_2(1270)$ (model 1), the $f_2(1270)$ mass and width are fixed to the world-average values, while
the $f_0(500)$ parameters are allowed to vary. This model is fitted both simultaneously
to the $\Dsonep$ and $\Dsoneprp$ samples and separately to each sample.
An extended model adds an $f_0(980)$ contribution with parameters fixed
to those used in the double-charm \Bdecay analysis (model 2), and a
further variant replaces the $f_0(500)$ state by a QMI $S$-wave parametrisation, 
still combined with a fixed $f_2(1270)$ resonance (model 3).

The fits with $f_0(500)$ and $f_2(1270)$ states alone do not yield an acceptable
description in the combined fit to both $\D_{s1}$ states, although they
can describe each decay separately with significantly different
$f_0(500)$ parameters. Including the $f_0(980)$ component produces a good combined
fit, but with a strongly interfering $f_0(500)$--$f_0(980)$
system that leads to an unphysical $\pip\pim$ $S$--wave and markedly
different $f_0(500)$/$f_0(980)$ compositions in the two decays. In particular,
the $\Dsoneprp$ amplitude becomes dominated by the $f_0(980)$ contribution,
while the $f_2(1270)$ fraction is larger in $\Dsonep$ than in
$\Dsoneprp$, despite the latter having more available phase space. The
QMI $S$-wave model, in which the $\pip\pim$ $S$--wave is described by a generic spline constrained to be common to both decays, fails to
provide an adequate description in terms of likelihood and binned
$\chi^2$.

\subsection{Models with \texorpdfstring{\boldmath{$\Ds\pipm$}}{Dspi} resonances}

\begin{table}[tb]
    \centering
    \caption{Summary of amplitude models considered in the analysis, with the corresponding negative log-likelihood differences relative to the baseline model (highlighted in bold) and binned $\chi^2$ for the simultaneous fits to the $\Dsonep$ and \Dsdecaypr samples with 54 to 56 effective degrees of freedom, respectively. $N_{\rm par}$ is the number of floating parameters in the fits. }
    \label{tab:model_summary}
    \resizebox{\textwidth}{!}{
    \begin{tabular}{lrrrr}
\toprule
Model & \hspace{-15mm}$-\Delta\log\mathcal{L}$ & \hspace{-2mm}$N_{\rm par}$ & $\chi^2_{2460}$ & $\chi^2_{2536}$ \\
\midrule
\phantom{0}1. $f_0(500)$, $f_2(1270)$ & $73.7$ & 6 & 146.5 & 114.5 \\
\phantom{0}2. $f_0(500)$, $f_0(980)$, $f_2(1270)$ & $-2.5$ & 10 & 61.9 & 73.8 \\
\phantom{0}3. QMI $\pi\pi$, $f_2(1270)$ & $44.8$ & 12 & 106.2 & 109.1 \\
\midrule
\phantom{0}4. $f_0(500)$, RBW $T^*_{c\bar{s}0}$ & $-3.1$ & 8 & 64.6 & 67.8 \\
\phantom{0}5. $f_0(500)$, RBW $T^*_{c\bar{s}1}$ & $-7.3$ & 12 & 61.5 & 62.7 \\
\midrule
\boldmath{\phantom{0}6. $f_0(500)$}, {\bf SLA} \boldmath{$T^*_{c\bar{s}0}$} & $0$ & 8 & 67.0 & 71.1 \\
\phantom{0}7. $f_0(500)$, SLA $T^*_{c\bar{s}0}$, no isosymmetry & $-2.7$ & 12 & 63.1 & 69.8 \\
\phantom{0}8. $f_0(500)$, SLA $T^*_{c\bar{s}0}$, $\gamma_2\neq 0$ & $-2.7$ & 9 & 64.0 & 68.5 \\
\midrule
\phantom{0}9. $f_0(500)$, Tri$(D^*K)$ & $4.0$ & 6 & 77.4 & 69.2 \\
10. $f_0(500)$, Tri$(DK^*)$ & $12.7$ & 6 & 89.7 & 71.2 \\
11. $f_0(500)$, Tri$(D^*K)$ in $\Dsonep$, Tri$(DK^*)$ in $\Dsoneprp$ & $4.9$ & 6 & 75.0 & 69.7 \\
\midrule
12. $f_0(500)$, SLA $T^*_{c\bar{s}0}$ only in $\Dsonep$ & $13.1$ & 6 & 69.0 & 82.6 \\
13. QMI $\pi\pi$, SLA $T^*_{c\bar{s}0}$ & $-8.3$ & 14 & 56.4 & 63.7 \\
14. QMI $\pi\pi$, SLA $T^*_{c\bar{s}0}$ only in $\Dsonep$ & $1.4$ & 12 & 68.5 & 67.7 \\
15. $f_0(500)$, $f_2(1270)$, SLA $T^*_{c\bar{s}0}$ & $-5.9$ & 12 & 62.4 & 67.2 \\
16. $f_0(500)$, $f_2(1270)$, SLA $T^*_{c\bar{s}0}$ only in $\Dsonep$ & $5.8$ & 10 & 67.7 & 75.7 \\
17. $f_0(500)$, $f_2(1270)$ in $\Dsoneprp$, SLA $T^*_{c\bar{s}0}$ in $\Dsonep$ & $6.0$ & 8 & 69.8 & 74.6 \\
\bottomrule
\end{tabular}

    }
\end{table}

Given the limitations of models containing only $\pip\pim$ resonances, amplitudes with explicit resonant contributions in the $\Dsp\pipm$ channels are
considered. The first of this family (model 4 in Table~\ref{tab:model_summary}) includes scalar Breit--Wigner resonances in the
$\Ds\pipm$ channels in addition to a scalar $\pip\pim$ component, as
described in Sec.~\ref{sec:helicity_amplitudes}. The free parameters are
the couplings of the isospin-conjugate $T_{c\bar s0}^{*0}$ and
$T_{c\bar s0}^{*++}$ states which are independent for the \Dsonep and \Dsoneprp decays, 
as well as the masses and widths of the $T^*_{c\bar s0}$ and $f_0(500)$ states.

Profile-likelihood scans of the $T^*_{c\bar s0}$ mass and width,
using either both $\D_{s1}$ samples or only the $\Dsonep$ sample, 
reveal a single preferred solution in which the $T^*_{c\bar s0}$ mass lies above the upper kinematic boundary of the
$\Dsp\pipm$ spectrum, at around 2500 to 2600\mev, and the width remains poorly constrained, in a range from 0 to 200\mev. Several
variants, differing in the treatment of the $\pip\pim$ $S$--wave and in
whether both decays or only the $\Dsonep$ sample are included, all provide
acceptable fit quality and consistent $T^*_{c\bar s0}$ masses. In all
cases, the sum of fit fractions for the three largely overlapping
components significantly exceeds $100\%$, indicating strong destructive
interference between the $\pip\pim$ and $\Dsp\pipm$ amplitudes, 
which is natural given their similar scalar structure and a limited phase space. 

Another model (model 5) introduces vector $T^*_{c\bar s1}$ Breit--Wigner resonances in the
$\Ds\pipm$ channels. Likelihood scans in the $T^*_{c\bar s1}$ mass--width plane
favour a solution with a mass near or slightly above the kinematic boundary 
around 2600\mev and either a very small or a very large width. Additional local minima are found for masses below the $\Ds\pipm$ threshold. The models with vector $T^*_{c\bar s1}$ states generally achieve a slightly better fit quality than the models based on scalar $T^*_{c\bar s0}$ contributions, 
but at the cost of extra degrees of freedom associated with the
$D$-wave components. However, the vector $T^*_{c\bar s1}$ solutions are strongly 
disfavoured by the double-charm $B$-decay data~\cite{LHCb-PAPER-2024-033},
where the full multidimensional kinematics is available, and thus are not considered further.

\subsection{\texorpdfstring{\boldmath{$K$}}{K}-matrix description of \texorpdfstring{\boldmath{$\Ds\pi$}}{Dspi} scattering}

To model the bound state with the mass near the $\D K$ threshold, a two-channel $K$-matrix formalism in the scattering-length approximation is employed, as described in Sec.~\ref{sec:lineshapes}. In this approach, the $\Ds\pi$ and $DK$ channels are coupled through a common pole whose location is parametrised by the $K$-matrix parameters $\gamma$ and $\beta$.

\begin{figure}[p]
    \centering
    \includegraphics[width=0.4\linewidth]{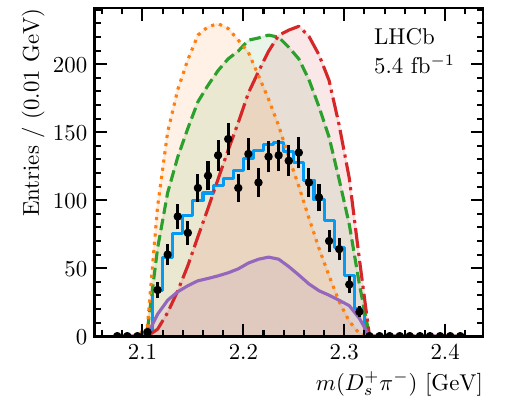}
    \put(-40,40){(a)}
    \includegraphics[width=0.4\linewidth]{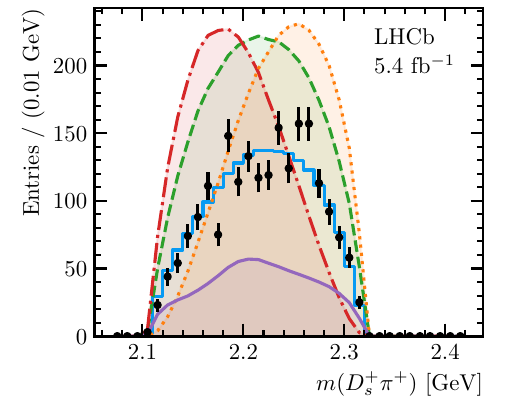}
    \put(-40,40){(b)}
    
    \includegraphics[width=0.4\linewidth]{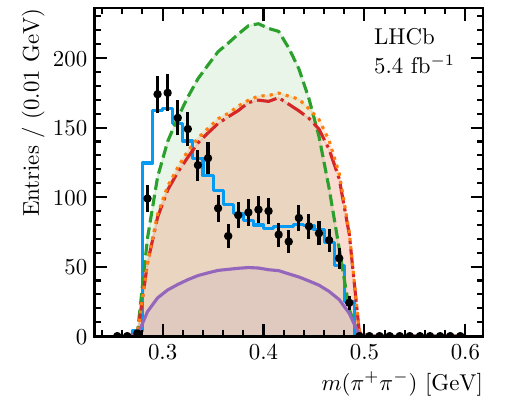}
    \put(-40,40){(c)}

    \includegraphics[width=0.4\linewidth]{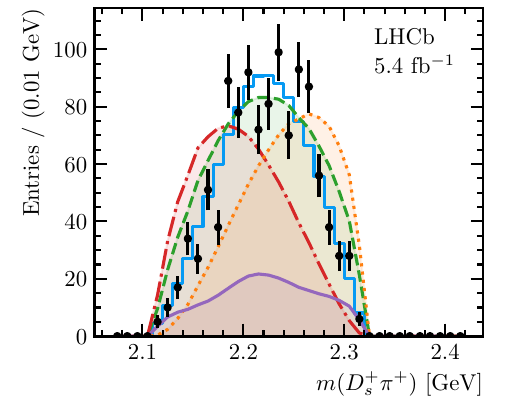}
    \put(-40,40){(d)}
    \includegraphics[width=0.4\linewidth]{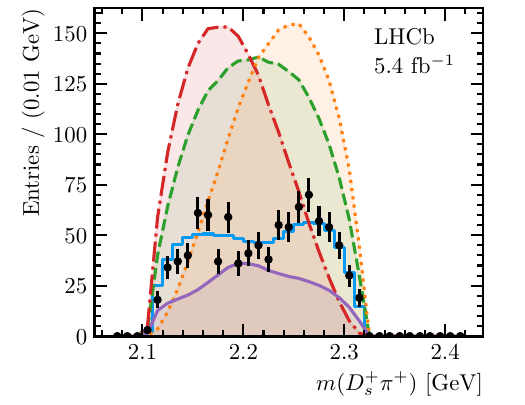}
    \put(-40,40){(e)}

    \includegraphics[width=0.4\linewidth]{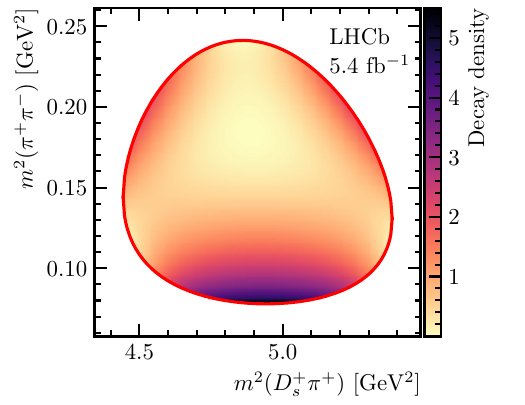}
    \put(-140,120){(f)}
    \includegraphics[width=0.4\linewidth]{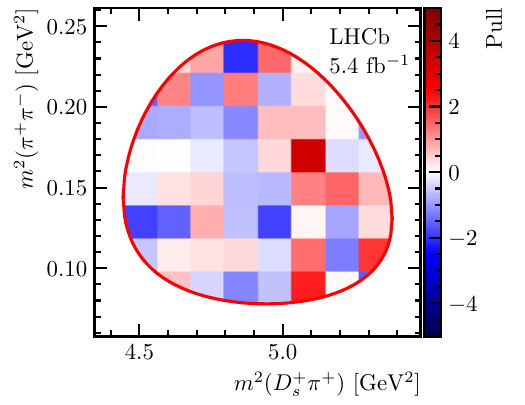}
    \put(-140,120){(g)}

    \includegraphics[width=0.7\linewidth]{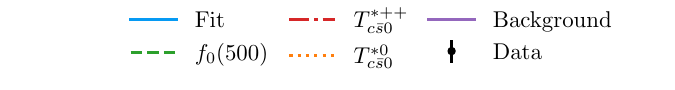}
    \vspace{-2mm}

    \caption{Results of the baseline amplitude fit to the \Dsdecay sample: (a,b) $m(\Dsp\pipm)$ projections of the fitted Dalitz-plot density with data overlaid, (c) $m(\pip\pim)$ projection, (d) $m(\Dsp\pip)$ projection of the slice with $m(\pip\pim)<0.36\gev$ and (e) $m(\pip\pim)>0.36\gev$, 
    (f) fitted density and (g) binned pull distribution. }
    \label{fig:ampl1}
\end{figure}

\begin{figure}[p]
    \centering
    \includegraphics[width=0.4\linewidth]{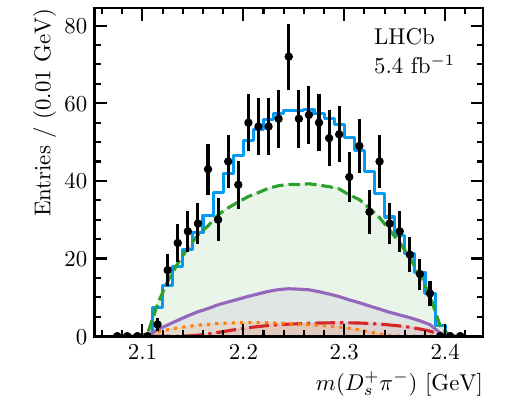}
    \put(-140,120){(a)}
    \includegraphics[width=0.4\linewidth]{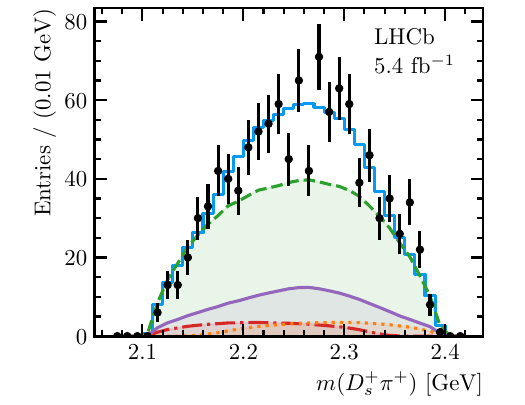}
    \put(-140,120){(b)}
    
    \includegraphics[width=0.4\linewidth]{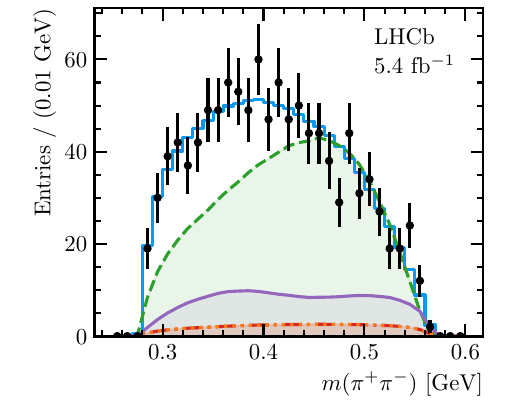}
    \put(-140,120){(c)}

    \includegraphics[width=0.4\linewidth]{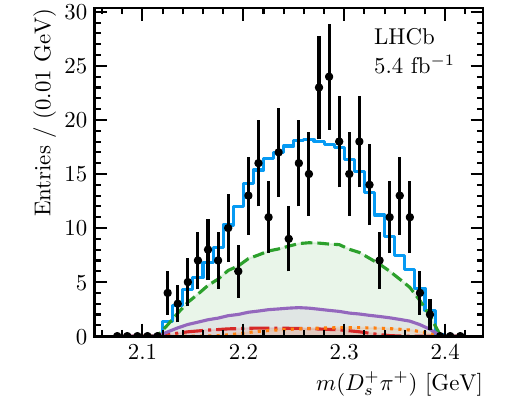}
    \put(-140,120){(d)}
    \includegraphics[width=0.4\linewidth]{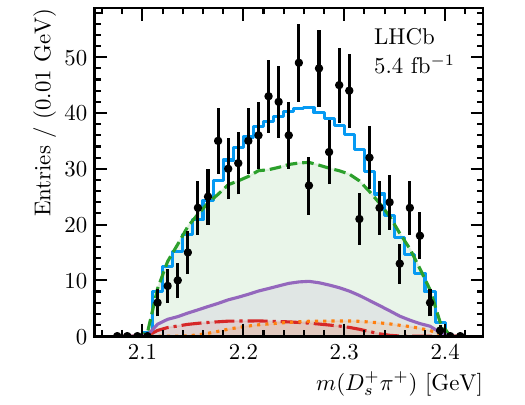}
    \put(-140,120){(e)}

    \includegraphics[width=0.4\linewidth]{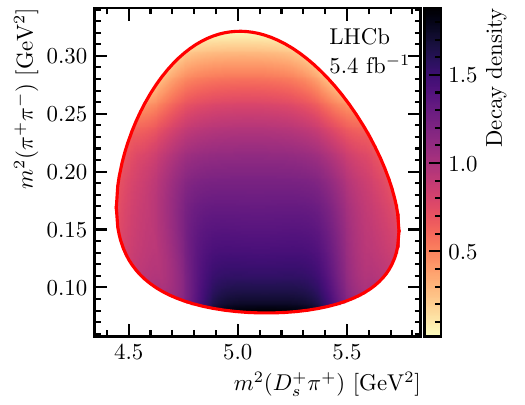}
    \put(-140,120){(f)}
    \includegraphics[width=0.4\linewidth]{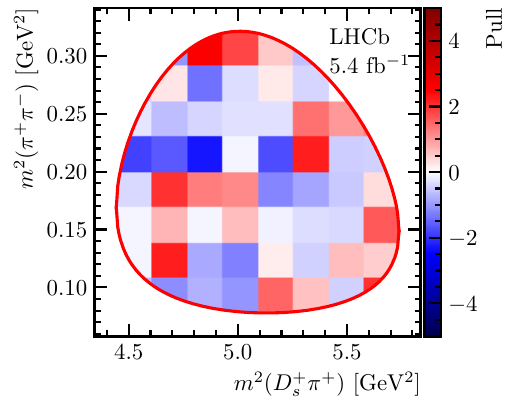}
    \put(-140,120){(g)}

    \includegraphics[width=0.7\linewidth]{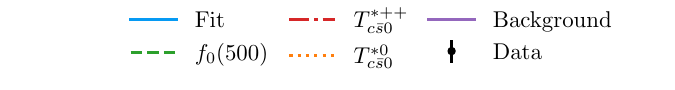}
    \vspace{-2mm}

    \caption{Results of the baseline amplitude fit to the \Dsdecaypr sample: (a,b) $m(\Dsp\pipm)$ projections of the fitted Dalitz-plot density with data overlaid, (c) $m(\pip\pim)$ projection, (d) $m(\Dsp\pip)$ projection of the slice with $m(\pip\pim)<0.36\gev$ and (e) $m(\pip\pim)>0.36\gev$, 
    (f) fitted density and (g) binned pull distribution. }
    \label{fig:ampl2}
\end{figure}

\begin{figure}
    \centering
    \includegraphics[width=0.48\linewidth]{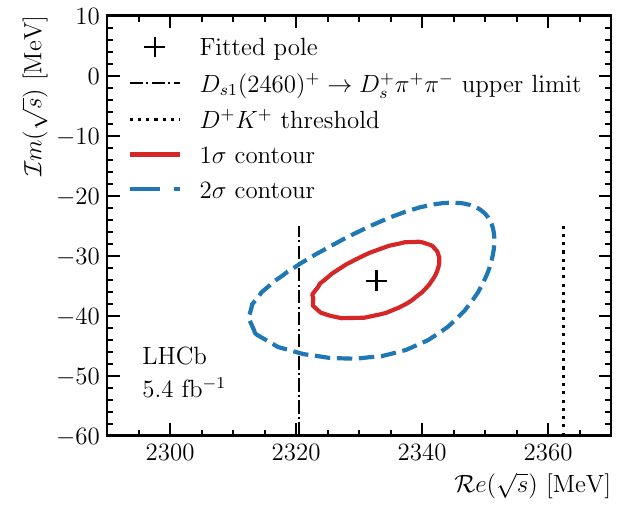}
    \put(-42,45){(a)}
    \includegraphics[width=0.48\linewidth]{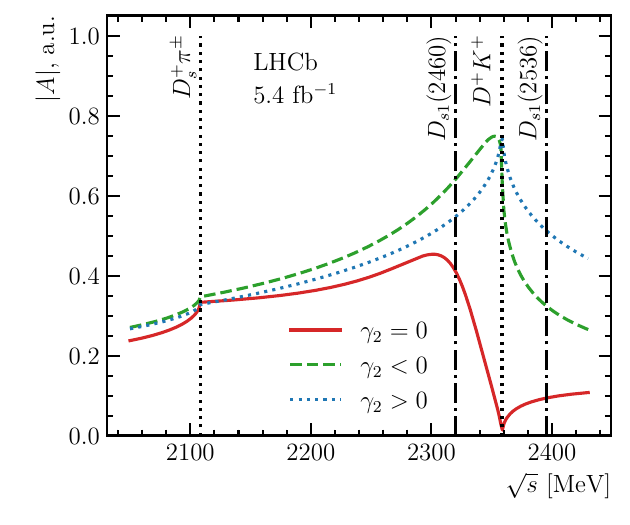}
    \put(-172,45){(b)}

    \caption{(a) Constraints on the pole position in the complex $\sqrt{s}$
    plane obtained from the baseline $K$-matrix fit with $\gamma_2=0$ to
    the combined $\Dsonep$ and $\Dsoneprp$ samples. The contours
    correspond to one- and two-standard-deviation uncertainties.
    (b) Absolute values of the $\Ds\pi$ amplitude from the fit results
    with $\gamma_2=0$ (solid line) and for the two solutions with
    non-zero $\gamma_2$ (dashed and dotted lines). The vertical lines
    show the $\Ds\pipm$ and $\Dp\Kp$ thresholds (dotted) and the upper 
    kinematic limits for the $\Dsonep$ and $\Dsoneprp$ decays to \Dspp (dash-dotted).
    }
    \label{fig:pole_contours}
\end{figure}

A few variants of this model are studied. In the simplest one (model 6), the parameter
$\gamma_2$ responsible for the direct $\Ds\pi$ interactions is fixed to zero,
corresponding to the scenario in which the $\Ds\pi$ interaction is
entirely driven by coupling to the $DK$ channel. The combined fit to the
$\Dsonep$ and $\Dsoneprp$ samples using this model provides a good
description of the data, as illustrated in Figs.~\ref{fig:ampl1} and \ref{fig:ampl2}. 
A scan in the $(\gamma,\beta)$ plane shows a single well-defined minimum with negative $\gamma$, characteristic of a bound $DK$ state. The corresponding constraints of the pole position in the complex $\sqrt{s}$ plane are shown in Fig.~\ref{fig:pole_contours}(a). 
The fitted parameters and derived pole position are summarised in Table~\ref{tab:ampl_results}. A fit to the $\Dsonep$ sample alone yields very similar $K$-matrix parameters. Being the minimal physically well-motivated model that gives a reasonable description of data, model 6 is taken as the baseline, as in Ref.~\cite{LHCb-PAPER-2024-033}. 

To probe possible isospin-breaking effects in the $T^*_{c\bar s0}$ system, an
additional fit is performed in which the complex couplings of the
$\Dsp\pip$ and $\Dsp\pim$ amplitudes are allowed to vary independently, 
while the $K$-matrix parameters $\gamma$ and $\beta$ are kept common (model 7). 
The likelihood of this fit shows no significant improvement with respect to the
isospin-symmetric baseline, which indicates no evidence for isospin
asymmetry in $T^*_{c\bar s0}$ production at the present level of
precision.

Allowing for a nonzero $\gamma_2$ leads to a three-parameter
description that is not fully constrained by the present data (model 8). The
likelihood landscape in this case contains extended regions of nearly 
equivalent solutions for correlated $\gamma$ and $\gamma_2$, both
negative and positive, and the baseline $\gamma_2=0$ solution
lies within the negative-$\gamma$ region. The improvement in likelihood
and binned $\chi^2$ with respect to the simpler model is marginal.

The absolute values of the $\Ds\pi$ amplitudes for the
different $K$-matrix solutions are shown in
Fig.~\ref{fig:pole_contours}(b). The model with $\gamma_2=0$
predicts a characteristic minimum of the $\Dsp\pi$ amplitude around the $DK$
threshold. Since this threshold lies within the phase space of the
$\Dsoneprp\to\Dsp\pip\pim$ decay, larger data samples of this mode
should clarify the structure of the amplitude in the $DK$ threshold region.

\subsection{Triangle-diagram \boldmath\texorpdfstring{$\Dsp\pipm$}{Ds pi} lineshapes}
\label{sec:triangle_lineshapes}

The ``cusp'' lineshapes in $\Dsp\pipm$ channels expected from the triangle diagrams are tested in the fit to data (models 9--11). Of all the models tried, the $\Dstar K$ triangle diagram, denoted as ``Tri($\Dstar K$)'' in Table~\ref{tab:model_summary}, provides the best fit quality for the \Dsonep decay. The \Dsoneprp decay is instead insensitive to the type of triangle diagram, with the $\D\Kstar$ diagram, ``Tri($\D\Kstar$)'', providing very similar fit results. The $\Dstar K$ triangle in the \Dsonep decay yields 4.0 units higher negative log likelihood for two fewer degrees of freedom compared to the baseline model. 
However, the binned $\chi^2$ value for the triangle amplitude is worse in the $\Dsonep$ sample, and, most strikingly, the parameters of the $\pip\pim$ $S$--wave differ significantly from those obtained in the baseline fit, and from those expected from other measurements. For example, the $f_0(500)$ mass reaches the limit at around 1.0\gev, which suggests that the data prefer the solution without phase rotation as a function of the $\pip\pim$ invariant mass. 
One can therefore conclude that the triangle diagram model used in the fit is disfavoured compared to the baseline. 

\subsection{Systematic uncertainties}
\label{sec:syst}

The fitted parameters of the baseline amplitude model and their statistical
and systematic uncertainties are summarised in Table~\ref{tab:syst}.
The fit procedure is validated with pseudoexperiments, and small corrections 
are applied to the fitted values and statistical uncertainties to account for observed biases. 
Systematic uncertainties are evaluated by varying the ingredients of the fitting model and repeating
the full fit as described below. 

\renewcommand{\arraystretch}{1.05}
\begin{table}
\caption{Systematic uncertainties on the parameters of the baseline model.}
\label{tab:syst}
\centering
\resizebox{\textwidth}{!}{\begin{tabular}{l|rr|rr|rr|rr}
\toprule
          &          &         & \multicolumn{2}{c|}{$f_0(500)$} & \multicolumn{2}{c|}{$a^{(2460)}$} & \multicolumn{2}{c}{$a^{(2536)}$}\\
Parameter & $\gamma$ & $\beta$ & $m$ & $\Gamma$ & \Real & \Imag & \Real & \Imag \\
          &          &         & \multicolumn{2}{c|}{[MeV]} & \multicolumn{4}{c}{$(\times 10^{-4})$} \\
\midrule
Fit value & $-298$ & 156 & 528 & 268 & $-83$ & 230 & $-90$ & $-5$ \\
Statistical uncertainty & ${}_{-35}^{+25}$ & ${}_{-16}^{+20}$ & ${}_{-30}^{+50}$ & ${}_{-28}^{+36}$ & ${}_{-41}^{+36}$ & ${}_{-50}^{+51}$ & ${}_{-26}^{+22}$ & ${}_{-37}^{+26}$ \\
\midrule
Efficiency parametrisation & 11 & 4 & 7 & 4 & 3 & 10 & 4 & 4 \\
Efficiency, simulation sample size &  7 & 2 & 5 & 8 & 6 & 11 & 7 & 4 \\
Efficiency, simulation corrections & 12 & 2 & 8 & 7 & 2 & 12 & 7 & 3 \\
Efficiency, $B$ production & 5 & 1 & 1 & 7 & 1 & 13 & 5 & 1 \\
Efficiency, $D_{s1}\overline{D}$ composition & 9 & 2 & 5 & 5 & 1 & 12 & 8 & 4 \\
Background parametrisation & 7 & 11 & 2 & 2 & 15 & 11 & 2 & 4 \\
Background, statistical & 13 & 7 & 19 & 11 & 18 & 25 & 9 & 7 \\
Background fraction & 4 & 2 & 13 & 13 & 11 & 38 & 9 & 4 \\
Normalisation & $<1$ & $<1$ & $<1$ & 2 & 1 & 2 & 1 & $<1$ \\
$\D_{s1}$ polarisation & 4 & 5 & $<1$ & 1 & 3 & 8 & $<1$ & $<1$ \\
Blatt--Weisskopf radius & 12 & 3 & 22 & 18 & 23 & 14 & 4 & 8 \\
Nonresonant admixture & 3 & 9 & 22 & 10 & 22 & 7 & $-$ & $-$ \\
\midrule
Total systematic & 29 & 17 & 41 & 31 & 42 & 57 & 19 & 14 \\
\bottomrule
\end{tabular}
}
\end{table}
\renewcommand{\arraystretch}{1}

Several sources of systematic uncertainty are related to the
efficiency description discussed in Sec.~\ref{sec:efficiency}.
The uncertainty related to the parametrisation of the Dalitz-plot
efficiency is estimated by replacing the nominal third-order polynomial with
a second-order one or an alternative description based on
neural networks, similar to those used for the background
density. The largest deviation from the baseline result is taken as the 
corresponding uncertainty. 

The effect of the limited size of the simulated sample 
is evaluated by constructing multiple bootstrap replicas of the samples
used to determine the efficiency shape, and the standard deviation of the
refitted parameters over these replicas is assigned as the corresponding
systematic uncertainty. 

Corrections using the data are applied to the simulated sample to account for 
imperfect modelling of the PID performance, tracking efficiency and trigger response.
The impact of the PID calibration is evaluated by varying the kernel width used in the
PID response correction~\cite{LHCb-DP-2018-001} applied to simulation and
recomputing the efficiency profile. The effect of corrections to the tracking 
and trigger efficiencies is estimated by using the uncorrected efficiencies 
from simulation to compute alternative efficiency profiles. The resulting parameter shifts
are assigned as systematic uncertainties.

A further contribution arises from the modelling
of the initial $B$-meson production kinematics. It is estimated by matching 
simulated \Ddecaypr control samples to the background-subtracted~\cite{Pivk:2004ty} data 
in intervals of \pt and $\eta$ of the visible $\Dsone\mun$ combination,
and computing the alternative efficiency profile after applying these corrections. 
The fit is repeated with the modified efficiency, and the resulting
parameter shifts are taken as systematic uncertainties.

The composition of the $\B\to \Dsonep X$ sample used to determine the
efficiency is another potential source of bias. Its effect is estimated
by replacing the nominal mixture of \Bdecay decays by pure $\B\to\Dsonep\Db$ or
$\B\to\Dsonep\Dstarb$ samples, or by adding an admixture of the
semileptonic mode \mbox{$\Bs\to\Dsonep\mun\neumb$}. 

Several systematic effects are associated with the background
description. The effect of its Dalitz-plot density parametrisation is probed by modifying
the configuration of the neural-network parametrisation, changing both
the network architecture and the $L_2$ regularisation strength. The
maximum observed deviation with respect to the baseline results is taken
as the systematic uncertainty. The finite size of the sideband samples
used to determine the background shapes is again treated with
bootstrapping, and the spread of the refitted parameters is assigned as
the corresponding uncertainty. The uncertainty in the
background fraction in the signal region is evaluated by varying it
within the uncertainty of the corresponding invariant-mass fits. 

The numerical integration used to compute the normalisation of the signal and background probability densities is
varied by changing the size of the uniform phase-space grid from the nominal
$300\times 300$ points to $1000\times 1000$; the effect is found to be negligible compared to other sources.

The formalism used in the amplitude fit assumes that $\Dsone$ mesons are
produced unpolarised. However, nonuniform selection efficiency over the kinematic
variables describing the $\Dsone$ production may induce an effective polarisation.
This effect is studied by calculating the polarisation density matrix from the parametrisation of the efficiency as a function of the $\Dsonep$ helicity angles in simulated
$\B\to\Dsonep\Db$ decays and using it in the amplitude fit. 
The resulting parameter shifts are assigned as systematic uncertainties.

The dependence on the assumed Blatt--Weisskopf barrier
radius entering the Breit--Wigner lineshapes is assessed by varying the
radius between $3$ and $6\gev^{-1}$ around the nominal value
$d=4\gev^{-1}$, with the maximal deviation assigned as the corresponding
uncertainty. 

A further model-related contribution accounts for a possible coherent nonresonant admixture in the
\Dsdecay amplitude, for instance, from
double-charm \mbox{$\B\to D_1\Dsp$}, \mbox{$D_1\to D\pip\pim$} decays whose
\Dspp invariant mass falls near the $\Dsonep$ peak. The
procedure used to estimate this effect is described in
Appendix~\ref{sec:app_nonres}. For the \Dsdecaypr
channel, which is dominated by semileptonic $B$ decays, any interference
with amplitudes of different quantum numbers is expected to cancel, and
no additional uncertainty from nonresonant contributions is assigned.

\subsection{Amplitude fit results and discussion}

The results of the baseline amplitude fit with the $K$-matrix
description of the $\Ds\pi$ amplitudes are summarised in Table~\ref{tab:ampl_results},
including statistical and systematic uncertainties. The parameters of the $K$-matrix
$\gamma$ and $\beta$, the mass and width of the $f_0(500)$ resonance, the fit fractions
of the different amplitude components in \Dsonep and \Dsoneprp decays, 
and the position of the $\Ds\pi$ pole in the complex plane $\sqrt{s_0}$ are given. The parameters of both the $\Dsp\pipm$ amplitude and the $f_0(500)$ state are in good agreement with the results of Ref.~\cite{LHCb-PAPER-2024-033}, where the same model was used. 

\renewcommand{\arraystretch}{1.1}
\begin{table}[tb]
    \centering
    \caption{Results of the baseline fit to the $\Dsonep$ and $\Dsoneprp\to\Dsp\pip\pim$ samples, fit fractions of the components, and the position of the $\Ds\pi$ pole. The fit fractions of the $f_0(500)$ and $T^*_{c\bar{s}0}$ components calculated in the \Dsonep and \Dsoneprp decays are denoted by the subscript ``2460'' and ``2536'', respectively. }
    \label{tab:ampl_results}
    \begin{tabular}{lr}
\toprule
Parameter & Value \\
\midrule
$\gamma$ & $-298\,{}_{-35}^{+25}\pm 29$ \\
$\beta$ & $156\,{}_{-16}^{+20}\pm 17$ \\
$m(f_0(500))$ [MeV] & $528\,{}_{-30}^{+50}\pm 41$ \\
$\Gamma(f_0(500))$ [MeV] & $268\,{}_{-28}^{+36}\pm 31$ \\
\midrule
$\mathcal{F}(f_0(500))_{2460}$ (\%) & $293\pm \phantom{.}53\pm 113$ \\
$\mathcal{F}(T^*_{c\bar{s}0})_{2460}$ (\%) & $257\pm \phantom{.}31\pm \phantom{0}62$ \\
$\mathcal{F}(f_0(500))_{2536}$ (\%) & $102\pm \phantom{.}23\pm \phantom{00}9$ \\
$\mathcal{F}(T^*_{c\bar{s}0})_{2536}$ (\%) & $6.6\pm 2.1\pm \,2.8$ \\
\midrule
$\,\Real(\sqrt{s_0})$ [MeV] & $2333\pm 9\pm 11$ \\
$\Imag(\sqrt{s_0})$ [MeV] & $-34\pm 6\pm 7\phantom{0}$ \\
\bottomrule
\end{tabular}

\end{table}
\renewcommand{\arraystretch}{1}

Several additional amplitude compositions are investigated to test the
robustness of the conclusions that the $\Dsp\pipm$ amplitudes are required
in both the $\Dsonep$ and $\Dsoneprp$ decays. These
variants, together with the main models discussed above, are
summarised at the bottom of Table~\ref{tab:model_summary}. 

Removing the SLA $\Ds\pipm$ contributions from the
$\Dsoneprp$ amplitude in favour of a pure $\pip\pim$ $S$--wave (model 12)
significantly degrades the likelihood, indicating that an additional
component beyond the scalar $\pip\pim$ contribution is needed in this
decay. Using Wilks' theorem, the difference in the negative log-likelihood
values between the baseline model and this variant, $-2\Delta\ln\mathcal{L}=26.2$, 
with two fewer degrees of freedom, yields the statistical significance of 
the $\Ds\pi$ component in the $\Dsoneprp$ decay of $4.7$ standard deviations.
A slightly lower significance of about $4\sigma$ is obtained if the $\pip\pim$ 
$S$--wave is 
described by a more flexible QMI parametrisation (models 13, 14). Variants in which the tensor $f_2(1270)$ 
contribution is added together or instead of the SLA $\Ds\pi$ amplitude (models 15--17)
also indicate that the $\Ds\pi$ components are needed for a good description of the $\Dsoneprp$ decay
with similar significance.  

Overall, the studies performed with the different amplitude models
favour scenarios in which the $\Dsonep$ decay contains a sizeable
near-threshold $\Ds\pi$ component consistent with a $DK$ molecular
configuration, while the $\Dsoneprp$ decay is less strongly affected
but still disfavourable to a description in terms of a single
$\pip\pim$ $S$-wave amplitude.

\section{Branching fraction and mass measurement}
\label{sec:branching}

The masses of the $\Dsonep$ and $\Dsoneprp$ states and the ratio of branching
fractions for the $\Dsoneprp$ decays to the \Dspp and $\Dp\Kp\pim$ final states
are determined using the signal yields obtained from the invariant-mass
fits. Taking the yields $N_{\rm sig}$ of the $\Dsoneprp$ decays into \Dspp
and $\Dp\Kp\pim$, and correcting them by the ratio of
efficiencies $\varepsilon_{DK\pi}/\varepsilon_{\D_s\pi\pi}$ for these two modes, together with the efficiency-correction factors $r_{DK\pi}$ and $r_{\D_s\pi\pi}$ that account for the nonuniformity of the efficiency and decay density over the phase space, the combination of branching fractions is obtained as
\begin{equation}
  \begin{split}
      R_1 \equiv & 
      \frac{\BR(\Dsoneprp\to\Dsp\pip\pim)\,\BR(\Dsp\to\Kp\Km\pip)}
           {\BR(\Dsoneprp\to\Dp\Kp\pim)\,\BR(\Dp\to\Km\pip\pip)} 
       \\
       = & \frac{N_{\rm sig}(\Dsoneprp\to\Dsp\pip\pim)}{N_{\rm sig}(\Dsoneprp\to\Dp\Kp\pim)} \cdot 
         \frac{\varepsilon_{DK\pi}}
              {\varepsilon_{\D_s\pi\pi}} 
         \cdot \frac{r_{DK\pi}}{r_{\D_s\pi\pi}} = 0.912\pm 0.046.
  \end{split}\nonumber
\end{equation}
Here $\varepsilon_{DK\pi,\D_s\pi\pi}$ are obtained from the simulated samples
generated with uniform phase space, while the factors $r_{DK\pi,\D_s\pi\pi}$ provide 
percent-level corrections to account for the correlation between the efficiency profile 
and the decay density across the Dalitz plot:
\begin{equation}
    r_i = \frac{\int \varepsilon_i(\Omega)\,p^{(i)}_{\rm sig}(\Omega)\,\mathrm{d}\Omega}
             {\int \varepsilon_i(\Omega)\,\mathrm{d}\Omega \cdot \int p^{(i)}_{\rm sig}(\Omega)\,\mathrm{d}\Omega}. 
\end{equation}
The efficiency profile and the signal density for the \Ddecaypr decay
are described in Appendix~\ref{sec:app_dkpi}. 

Taking into account the ratio of branching fractions for the charm decays~\cite{PDG2024}, 
the ratio of branching fractions for $\Dsoneprp$ decays into the \Dspp and $\Dp\Kp\pim$ final states is obtained as
\begin{equation}
    R_2 \equiv 
    \frac{\BR(\Dsoneprp\to\Dsp\pip\pim)}
         {\BR(\Dsoneprp\to\Dp\Kp\pim)} = 1.57\pm 0.08.\nonumber
\end{equation}
Finally, using the absolute branching fraction for the $\Dsoneprp\to\Dp\Kp\pim$
decay from Ref.~\cite{PDG2024}, $\BR(\Dsoneprp\to\Dp\Kp\pim) =
(10.0\pm 2.5)\times 10^{-3}$, the branching fraction is obtained as
\begin{equation}
    \BR \equiv \BR(\Dsoneprp\to\Dsp\pip\pim) = (1.57\pm 0.08)\%.\nonumber
\end{equation}
The uncertainties of the $R_1$, $R_2$ and $\BR$ values quoted above are
statistical only and do not include the uncertainties of the external inputs.

The masses of the $\Dsonep$ and $\Dsoneprp$ states are measured from the $\Dsp\pip\pim$ and $\Dp\Kp\pim$ invariant-mass distribution, respectively. 
The measurement of \Dsoneprp mass from the $\Dsp\pip\pim$ final state is
not used due to the larger statistical uncertainty and the potentially larger bias
from interference with nonresonant components, as discussed below.
The fit parameters of the data distributions are the shifts $\Delta m$ of the signal peaks with respect
to the parametrised simulated distributions. The simulation uses the
latest world-average values, $2459.5\mev$ for $\Dsonep$ and $2535.11\mev$ for
$\Dsoneprp$~\cite{PDG2024}. Taking these values into account and adding the
mass shifts from data fits, the masses of the two states are measured
to be
\[
  m(\Dsonep) = 2459.51\pm 0.14\mev,
\]
and
\[
  m(\Dsoneprp) = 2535.18\pm 0.04\mev,
\]
where the uncertainties are statistical only.

The systematic uncertainties, as well as the central values and
statistical uncertainties of the measured quantities $R_1$, $R_2$, $\BR$,
$m(\Dsonep)$, and $m(\Dsoneprp)$, are summarised in
Table~\ref{tab:branching_mass}. The main systematic contributions
originate from the finite size of the simulated samples used to
determine efficiencies and signal shapes, from the calibration of the
PID response that enters the branching-fraction measurements, and from
the modelling of the signal and background shapes in the invariant-mass
fits. Additional effects arise from the efficiency-correction factors
$r$, which account for correlations between the efficiency profile and
the decay density, but are found to be small, as well as from the limited
precision of external branching fractions used in the calculation of
$R_2$ and $\BR$, namely $\BR(\Dsp\to\Kp\Km\pip)$,
$\BR(\Dp\to\Km\pip\pip)$, and $\BR(\Dsoneprp\to\Dp\Kp\pim)$. For the mass measurements, further
uncertainties are associated with the momentum-scale calibration of the
tracking system, known to a relative precision of $3\times 10^{-4}$~\cite{LHCb-DP-2023-003}, and
with the knowledge of the $\Dp$ and $\Dsp$ masses that are constrained
to their world-average values $m(\Dp)=1869.66\pm 0.05\mev$ and $m(\Ds)=1968.35\pm 0.07\mev$~\cite{PDG2024} in the fits and thus propagate to the
extracted values of $m(\Dsonep)$ and $m(\Dsoneprp)$.

\begin{table}
    \centering
    \caption{Central values and statistical and systematic uncertainties
      for the measurement of the branching-fraction ratios $R_1$ and
      $R_2$, the branching fraction of the \Dsdecaypr
      decay $\BR$, and the masses of the $\Dsonep$ and $\Dsoneprp$
      states.}
    \begin{tabular}{lrrrrr}
\toprule
     Source of uncertainty &  \multicolumn{1}{c}{$R_1$} &  \multicolumn{1}{c}{$R_2$} &  \multicolumn{1}{c}{$\BR$} &  \multicolumn{1}{c}{$m(\Dsonep)$} &  \multicolumn{1}{c}{$m(\Dsoneprp)$} \\
                           &        &        &  \multicolumn{1}{c}{(\%)}  &  \multicolumn{1}{c}{[MeV]}        &  \multicolumn{1}{c}{[MeV]} \\
\midrule
             Central value &  0.912 &  1.569 &       1.569 &        2459.510 &        2535.182 \\
   Statistical uncertainty &  0.046 &  0.079 &       0.079 &           0.137 &           0.042 \\
\midrule
             Simulation sample size &  0.013 &  0.023 &       0.023 &           0.053 &           0.012 \\
             Simulation correction &  0.018 &  0.031 &       0.031 &               - &               - \\
Efficiency correction factors &  0.003 &  0.004 &    0.004 &               - &               - \\
              Signal shape &  0.025 &  0.042 &       0.042 &           0.036 &           0.005 \\
          Background shape &  0.034 &  0.058 &       0.058 &        $<0.001$ &           0.002 \\
          Branching ratios &      - &  0.035 &       0.394 &               - &               - \\
Momentum scale calibration &      - &      - &           - &           0.091 &           0.015 \\
          $D_{(s)}^+$ mass &      - &      - &           - &           0.070 &           0.050 \\
\midrule
   Total syst. uncertainty &  0.047 &  0.089 &       0.402 &           0.131 &           0.054 \\
\bottomrule
\end{tabular}

    \label{tab:branching_mass}
\end{table}

The invariant-mass fits are performed under the assumption that there is
no interference between the $\D_{s1}$ signals and nonresonant or broad
resonant $\Dsp\pip\pim$ or $\Dp\Kp\pim$ components. Such interference
could modify both the yields and the masses of the $\D_{s1}$ states. The
estimation of these effects is discussed in
Appendix~\ref{sec:app_nonres}, and the corresponding uncertainties are
summarised in Table~\ref{tab:nonres_syst}. The total relative
uncertainty on the ratios $R_1$, $R_2$ and $\BR$, arising from the
\Dsdecaypr and \Ddecaypr decay modes,
is therefore 10.6\%. These estimates
are quoted separately when reporting the branching-fraction and
mass results. Further studies of the
\Bdecay and \Bsdecaypr amplitudes could
provide tighter constraints on the coherent nonresonant amplitudes and
thus reduce this uncertainty.

\begin{table}
    \centering
    \caption{Uncertainties on the mass and yield measurements arising
      from a possible coherent nonresonant admixture.}
    \label{tab:nonres_syst}
    \begin{tabular}{lcc}
        \toprule
        Mode & Relative yield uncertainty & Mass uncertainty [MeV] \\
        \midrule
         $\Dsonep\to\Dsp\pip\pim$    & 2.8\% & 0.15 \\
         $\Dsoneprp\to\Dp\Kp\pim$    & 5.5\% & 0.16 \\
         $\Dsoneprp\to\Dsp\pip\pim$  & 9.1\% & 0.37 \\
         \bottomrule
    \end{tabular}
\end{table}

Finally, measured branching fractions are
\begin{align*}
      \frac{\BR(\Dsoneprp\to\Dsp\pip\pim)\,\BR(\Dsp\to\Kp\Km\pip)}
           {\BR(\Dsoneprp\to\Dp\Kp\pim)\,\BR(\Dp\to\Km\pip\pip)} &= 
      0.91\pm 0.05\pm 0.05\pm 0.10, \\
    \frac{\BR(\Dsoneprp\to\Dsp\pip\pim)}
         {\BR(\Dsoneprp\to\Dp\Kp\pim)} &= 1.57\pm 0.08\pm 0.09\pm 0.17,  \\
\BR(\Dsoneprp\to\Dsp\pip\pim) &= (1.57\pm 0.08\pm 0.40\pm 0.17)\%,\end{align*}
and the determined masses are
\begin{align*}
  m(\Dsonep) &= (2459.51\pm 0.14\pm 0.13\pm 0.15)\mev, \\
  m(\Dsoneprp) &= (2535.18\pm 0.04\pm 0.05\pm 0.16)\mev,
\end{align*}
where the first uncertainty is statistical, the second systematic, and
the third due to the possible coherent nonresonant admixture. 

As shown in Sec.~\ref{sec:massfit}, the upper limit on the $\Dsonep$ width is found to be 1.28\mev (1.38\mev) at 90\% (95\%) confidence level, which is significantly better than the current limit of 3.5\mev~\cite{BaBar:2006eep}. This value includes the dominant systematic uncertainties due to the resolution modelling and the uncertainty of the $\Dsoneprp$ width used as a constraint in the fit.

The branching fraction of the $\Dsoneprp\to\Dsp\pip\pim$ decay is measured here for the first time. 
It can be compared to the indirect determination from Ref.~\cite{Bondar:2023ibd}. However, the latter 
assumed that the $\Dsoneprp$ decay is saturated by the $\Dstar K$ decays, while the recent 
experimental results suggest that these modes account for only about 68\% of the total decay width~\cite{PDG2024}. 
Thus, the branching fraction obtained in Ref.~\cite{Bondar:2023ibd} needs to be scaled down accordingly, 
yielding \mbox{$\BR(\Dsoneprp\to\Dsp\pip\pim) = (1.7\pm 0.6)\%$}, which is in good agreement with the direct 
measurement reported here.

\section{Conclusion}
\label{sec:conclusion}
 
An amplitude analysis of the $\Dsonep$ and $\Dsoneprp$ decays to \Dspp is presented,
using samples tagged by an accompanying muon collected with the LHCb detector in proton-proton collisions at a centre-of-mass energy of 13\tev, corresponding to an integrated luminosity of 5.4\invfb.
Already at the level of the raw Dalitz plots, the two modes exhibit
visibly different structures, even though the parent states have the
same quantum numbers and very similar masses, implying that their decay
dynamics cannot be identical. This points to a difference in the
internal structure of the two states, such as a sizable molecular
$D^{*}K$ component in the $\Dsonep$ wave function that is absent or
strongly suppressed for the $\Dsoneprp$ state.

The results of this analysis strengthen the conclusions from the study of the \Dsonep decays produced in double-charm $B$ decays~\cite{LHCb-PAPER-2024-033}, where only the \Dsonep state is accessible. The description of \Dsdecay decay in the analysis of Ref.~\cite{LHCb-PAPER-2024-033} in terms of pure $\pip\pim$ dynamics was possible, but was shown to be unlikely due to the necessity of implausibly large $f_2(1270)$ and $f_0(980)$ admixture. In the current analysis, the models containing only scalar and tensor $\pip\pim$ resonances fail to provide a coherent description of both states. A good simultaneous description is obtained when allowing for additional
scalar contributions in the $\Dsp\pipm$ channels. In this framework, the
$\pip\pim$ $S$--wave is consistent with a common Breit--Wigner-like
$f_0(500)$ contribution for the decays of two resonances, while the $\Dsp\pipm$
component is much more pronounced in $\Dsonep$ than in $\Dsoneprp$ decays and
exhibits an enhancement near the $DK$ threshold. 

Parametrisation of the $\Dsp\pipm$ components with relativistic Breit--Wigner lineshapes yields good fit quality for the combined fit, however, the fitted mass of the resonance tends to be above the upper kinematic thresholds, and the width is poorly constrained. The dynamics of the $\Dsp\pipm$ channel are well described by a coupled-channel $K$-matrix model in which a near-threshold $DK$ bound
state couples to $\Ds\pi$. The preferred solution corresponds to a
pole located just below the $DK$ threshold, corresponding to a mass of
$2333\pm 9\stat\pm 11\syst\mev$ and a width of $68\pm 12\stat\pm
14\syst\mev$, consistent with results obtained from an independent sample of Ref.~\cite{LHCb-PAPER-2024-033}. No evidence for isospin breaking between the isospin partners 
in this system is found within the current precision. The description of the $\Dsp\pipm$ structures by triangle-loop amplitudes involving $\Dstar K$ or $\D\Kstar$ rescattering, using a simple model proposed in Ref.~\cite{LHCb-PAPER-2019-014}, is disfavoured, which suggests that phase rotation as a function of $\Dsp\pipm$ invariant mass is necessary to describe the data. 

In addition, the ratio of branching
fractions for $\Dsoneprp$ decays to the $\Dsp\pip\pim$ and
$\Dp\Kp\pim$ final states is measured for the first time, 
corresponding to ${1.57\pm 0.08\pm 0.09 \pm 0.17}$.
The masses of the $\Dsonep$ and $\Dsoneprp$ mesons are
determined, yielding ${2459.51\pm 0.14\pm 0.13 \pm 0.15\mev}$ and
${2535.18\pm 0.04\pm 0.05\pm 0.16\mev}$, respectively, 
where the first uncertainty is statistical, the second is systematic, and the third arises from the effects of possible nonresonant production of the 
$\Dsp\pip\pim$ and $\Dp\Kp\pim$ combinations in $\B$-meson decays. 
These measurements are consistent with, and more precise than, the
current world averages~\cite{PDG2024}. Finally, the upper limit on the $\Dsonep$ width 
is set to 1.38\mev at 95\% confidence level, which is the most stringent constraint to date.

\section*{Acknowledgements}
%
%
\noindent We acknowledge important input from Alex Bondar who initiated and contributed to all aspects of the analysis reported here. 

We express our gratitude to our colleagues in the CERN
accelerator departments for the excellent performance of the LHC. We
thank the technical and administrative staff at the LHCb
institutes.
We acknowledge support from CERN and from the national agencies:
ARC (Australia);
CAPES, CNPq, FAPERJ and FINEP (Brazil); 
MOST and NSFC (China); 
CNRS/IN2P3 and CEA (France);  
BMFTR, DFG and MPG (Germany);
NKFIH (Hungary);              
INFN (Italy); 
NWO (Netherlands); 
MNiSW and NCN (Poland); 
MEC/IFA (Romania); 
MICIU and AEI (Spain);
SNSF and SER (Switzerland); 
NASU (Ukraine); 
STFC (United Kingdom); 
DOE NP and NSF (USA).
We acknowledge the computing resources that are provided by ARDC (Australia), 
CBPF (Brazil),
CERN, 
IHEP and LZU (China),
IN2P3 (France), 
KIT and DESY (Germany), 
INFN (Italy), 
SURF (Netherlands),
Polish WLCG (Poland),
IFIN-HH (Romania), 
PIC (Spain), CSCS (Switzerland), 
GridPP (United Kingdom),
and NSF (USA).  
We are indebted to the communities behind the multiple open-source
software packages on which we depend.
Individual groups or members have received support from
RTP (Australia), 
FWO Odysseus grant G0ASD25N (Belgium), 
Key Research Program of Frontier Sciences of CAS, CAS PIFI, CAS CCEPP (China); 
Minciencias (Colombia);
EPLANET, Marie Sk\l{}odowska-Curie Actions, ERC and NextGenerationEU (European Union);
A*MIDEX, ANR, IPhU and Labex P2IO, and R\'{e}gion Auvergne-Rh\^{o}ne-Alpes (France);
Alexander-von-Humboldt Foundation (Germany);
ICSC (Italy); 
Severo Ochoa and Mar\'ia de Maeztu Units of Excellence, GVA, XuntaGal, GENCAT, InTalent-Inditex and Prog.~Atracci\'on Talento CM (Spain);
the Leverhulme Trust, the Royal Society and UKRI (United Kingdom).




\vspace{1.5\baselineskip}
{\noindent\normalfont\bfseries\Large Appendices}

\appendix

\section{Dalitz-plot distributions of \texorpdfstring{\mbox{\boldmath{$\Dsoneprp\to\Dp\Kp\pim$}}}{Ds1(2536) to D+ K+ pi-} decays}
\label{sec:app_dkpi}

The combined fit of the \Ddecaypr and \Dsdecaypr amplitudes 
could provide additional constraints on the parameters of the $T_{c\bar{s}0}$ amplitudes. However, 
since their contribution in the \Dsdecaypr decay
is small, 
and the much smaller phase space additionally suppresses the contribution of these amplitudes in the \Ddecaypr decay, 
the current data sample does not provide significant constraints. The Dalitz-plot analysis 
of the \Ddecaypr decay is nevertheless used to constrain the parameter $r_{DK\pi}$ needed to correct 
the ratio of branching fractions, as described in Sec.~\ref{sec:branching}.

\begin{figure}
    \centering
    \includegraphics[width=0.47\linewidth]{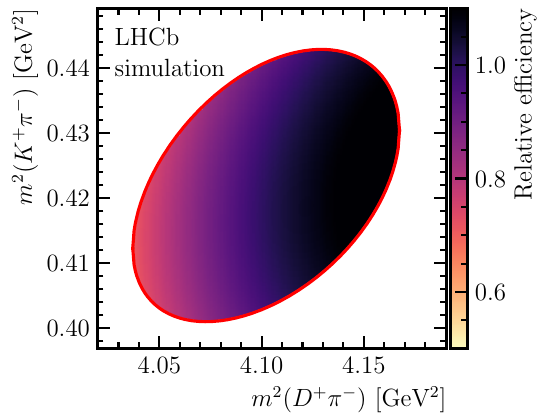}
    \caption{Relative efficiency profile across the Dalitz plots for $\Dsoneprp\to\Dp\Kp\pim$ decays.}
    \label{fig:efficiency_dkpi}
\end{figure}

The polynomial efficiency parametrisation for the \Ddecaypr decay is illustrated in Fig.~\ref{fig:efficiency_dkpi}. 
The parameterisation is similar to that used for the \mbox{$\Dsone\to\Dsp\pip\pim$} channels, and is based on the third-order polynomial in
the squared invariant masses $m^2(\Dp\pim)$ and $m^2(\Kp\pim)$ rescaled to the $(-1,1)$ range. 

\begin{figure}
    \centering
    \includegraphics[width=0.4\linewidth]{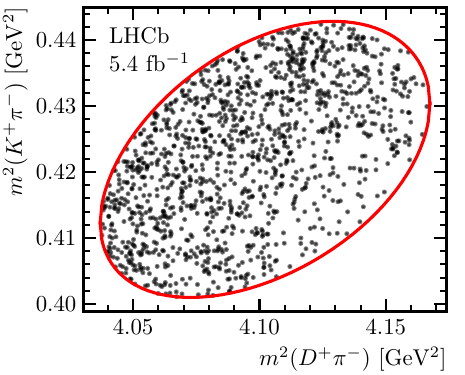}
    \put(-30,38){(a)}
    \includegraphics[width=0.4\linewidth]{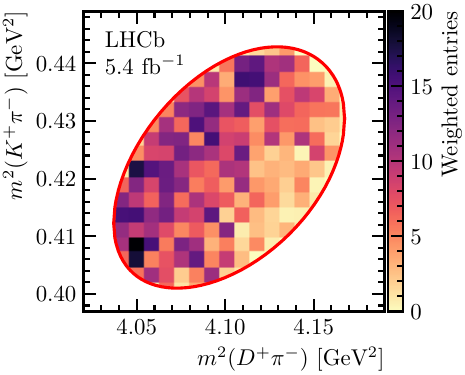}
    \put(-55,38){(b)}

    \includegraphics[width=0.4\linewidth]{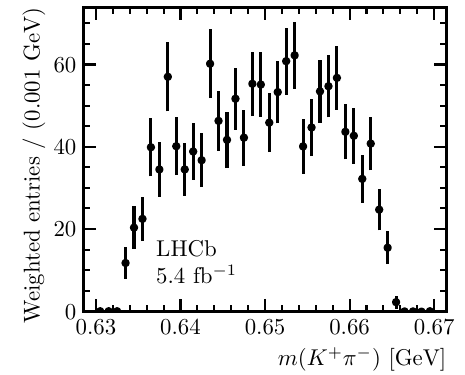}
    \put(-140,122){(c)}
    \includegraphics[width=0.4\linewidth]{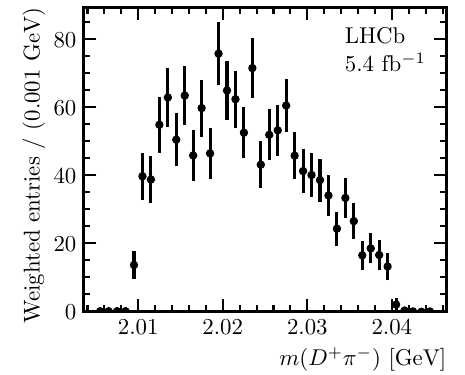}
    \put(-140,122){(d)}
    \caption{Distributions for the \Ddecaypr sample. (a) Raw Dalitz-plot distribution. Background-subtracted and efficiency-corrected distributions: (b) Dalitz plot, and (c) $m(\Kp\pim)$ and (d) $m(\Dp\pim)$ projections.}
    \label{fig:dkpi_dlz}
\end{figure}

The Dalitz-plot distribution of the \Ddecaypr decays and its projections are shown in Fig.~\ref{fig:dkpi_dlz}. The signal density $p_{\rm sig}(\Omega)$ used to compute the correction factor $r_{DK\pi}$ is obtained from an amplitude analysis of this sample, following the same formalism as described in 
Sec.~\ref{sec:formalism}. The amplitude model includes contributions from the scalar $\D_{0}^*(2300)^0$ resonance decaying to $\Dp\pim$, as well as scalar $K^*_0(700)^0$ and vector $K^*(892)^0$ resonances in the $\Kp\pim$ channel. The resonances are modelled by relativistic Breit--Wigner lineshapes with the parameters from Ref.~\cite{PDG2024}. Given the small phase space available in this decay, the exact parameter values and details of the amplitude model have a negligible effect on the efficiency-correction factor. 

\section{Nonresonant contributions to \texorpdfstring{\mbox{\boldmath{$\D_{s1}^+\to\Dsp\pip\pim$}}}{Ds1 to Ds+ pi+ pi-} decays}
\label{sec:app_nonres}

The reconstructed \Dspp and $\Dp\Kp\pim$ final states can receive coherent
contributions from amplitudes other than the narrow $\Dsonep$ and $\Dsoneprp$
resonances. Such contributions may originate from broad intermediate states in
four-body $B$ decays and can bias the measured masses, the branching
fractions, and the parameters of the Dalitz-plot amplitude model. In the following, these
additional contributions are referred to as ``nonresonant'', although in
practice they may arise from other resonant components. The production mechanisms
for $\Dsonep$ and $\Dsoneprp$ are different, and their possible nonresonant
admixtures are studied separately.

For the \Dsdecay decay, the nonresonant yield is constrained using data
on fully reconstructed double-charm $B\to\Dsp\pip\pim\Db^{(*)}$ decays~\cite{LHCb-PAPER-2024-033}.
After subtracting non-$B$ background using the \sPlot technique~\cite{Pivk:2004ty}, the $m(\Dsp\pip\pim)$
distributions are dominated by the $\Dsonep$ peak, with a small flat
contribution consistent with a few percent of the signal yield in
the $\pm10\mev$ window around the $\Dsonep$ mass. The pattern and size of
this excess are compatible with expectations from
$B\to\Dsp\D_1(2420)$, $\D_1(2420)\to(\Dzb\pip\pim)_\text{nonres}$ decays,
based on existing measurements of $B\to\D_1(2420)\Ds$~\cite{LHCb-PAPER-2024-001} and
$\D_1(2420)\to\D\pip\pim$~\cite{LHCb-PAPER-2011-016} branching fractions and on a
simulation of the four-body kinematics. The nonresonant $\Dsp\pip\pim$ contribution is therefore attributed to this amplitude and its effect is modelled under this
assumption.

The impact of this coherent admixture on the $\Dsonep$ mass and branching fraction is evaluated with a simplified model of the four-body decay
$B\to\Dsp\pip\pim\Db$ including both intermediate states. The decay density is
written as
\begin{equation}
    \mathrm{d}\Gamma = |\mathcal{M} + a_{\rm nr} e^{i\delta_{\rm nr}} \mathcal{M}'|^2\, \mathrm{d}\mathcal{P},
\end{equation}
where $d\mathcal{P}$ is the five-dimensional phase-space element,
$\mathcal{M}$ and $\mathcal{M}'$ denote the amplitudes with intermediate
$\Dsonep$ and $\D_1(2420)$ mesons, respectively, and $a_{\rm nr}$ and
$\delta_{\rm nr}$ are the relative magnitude and strong phase of the
$\D_1(2420)$ contribution. The $\Dsonep$ amplitude is taken in the form
\begin{equation}
    \mathcal{M} = \sum_{\nu} \mathcal{R}(m)\, d^{1}_{0,\nu}(\theta_1)\, e^{i\nu\phi_{23}}\, O_{\nu}(s_1,s_3),
\end{equation}
with $m = m(\Dsp\pip\pim)$ and a relativistic Breit--Wigner lineshape
$\mathcal{R}(m)$. The angle $\theta_1$ is the polar angle between the direction of 
the \Dsonep motion in the \B-meson rest frame and the direction of the $\pip\pim$ system in the 
rest frame of the \Dsonep state, while $\phi_{23}$ is the corresponding azimuthal angle of the $\pip\pim$ combination (see Ref.~\cite{JPAC:2019ufm}). 
The $\D_1(2420)$ amplitude $\mathcal{M}'$ has the same
structure with $\Dsp$ and $\Dzb$ interchanged and the corresponding mass and
width. The parameter $a_{\rm nr}$ is tuned such that the ratio of integrals of
the two densities in the $\Dsp\pip\pim$ invariant mass window matches the nonresonant
fraction observed in data, while $\delta_{\rm nr}$ is varied to maximise the
effect. The width of the $\Dsonep$ state is not measured and is estimated 
to be as large as $1.5\mev$~\cite{Bondar:2025qzg}; this value is used to obtain conservative
estimates of the systematic uncertainties.

Simulated $m(\Dsp\pip\pim)$ spectra are then smeared with the experimental
resolution and fitted with the nominal one-dimensional model that assumes a
single $\Dsonep$ resonance. The biases of the fitted yield and mass with respect to
the values without the $\D_1(2420)$ admixture are taken as systematic uncertainties, 
and equal $2.8\%$ and $0.15\mev$, respectively. The same model is used to 
estimate the potential bias in the Dalitz-plot
parameters extracted from the \Dsdecay amplitude analysis.
Samples are generated with and without the $D_1(2420)\to D\pip\pim$ admixture
for several values of $\delta_{\rm nr}$, and are fitted with the baseline
amplitude model that does not include the nonresonant component. For each fit
parameter, the maximum observed deviation is taken as the corresponding
systematic uncertainty; these contributions are reported in the
``Nonresonant admixture'' component of Table~\ref{tab:syst}.

Possible nonresonant contributions to the $\Dsoneprp$ channels are studied 
assuming their production in semileptonic $B\to\Dsoneprp\mun\neumb$ decays. For
\Ddecaypr, fits to the $m^2_{\rm miss}$ distributions in
the $m(\Dp\Kp\pim)$ signal and sideband regions reveal a sizeable coherent
nonresonant component under the $\Dsoneprp$ peak. The ratio of resonant and
nonresonant yields in the signal window is translated into an effective
two-component model in $m(\Dp\Kp\pim)$, given by the coherent sum of a
Breit--Wigner lineshape for $\Dsoneprp$ with a fixed natural width of
$0.92\mev$ and a linear nonresonant term, convolved with the detector
resolution. Varying the relative strong phase between these components leads to
a maximum mass shift of $\pm0.16\mev$ and a maximum change of $\pm5.5\%$ in the
visible peak yield; these are used as systematic uncertainties on the
$\Dsoneprp$ mass and on the \Ddecaypr yield entering the
branching-ratio determinations.

For \Dsdecaypr, the same procedure is applied using
fits to the $m^2_{\rm miss}$ distributions in bins of $m(\Dsp\pip\pim)$.
In this case, the background level is higher, and the nonresonant component is
less pronounced, but still appears to be non-negligible. The $m(\Dsp\pip\pim)$ distribution is modelled as the sum of contributions for $\Dsonep$, $\Dsoneprp$ and
$\D_{s2}^*(2573)^+$ resonances and a constant nonresonant term. Varying the
relative phase between the $\Dsoneprp$ and nonresonant components yields a
maximum mass shift of $\pm0.37\mev$ and a maximum change of $\pm9.1\%$ in the
visible \Dsdecaypr peak yield. These values are used as
systematic uncertainties on the corresponding mass and yield measurements and
enter the nonresonant-admixture uncertainties quoted in the main text.


\addcontentsline{toc}{section}{References}
\bibliographystyle{LHCb}
\bibliography{main,standard,LHCb-PAPER,LHCb-CONF,LHCb-DP,LHCb-TDR}

\newpage
\centerline
{\large\bf LHCb collaboration}
\begin
{flushleft}
\small
R.~Aaij$^{39}$\lhcborcid{0000-0003-0533-1952},
M.~Abdelfatah$^{71}$,
A.S.W.~Abdelmotteleb$^{59}$\lhcborcid{0000-0001-7905-0542},
C.~Abellan~Beteta$^{53}$\lhcborcid{0009-0009-0869-6798},
F.~Abudin\'en$^{61}$\lhcborcid{0000-0002-6737-3528},
T.~Ackernley$^{63}$\lhcborcid{0000-0002-5951-3498},
A.A.~Adefisoye$^{71}$\lhcborcid{0000-0003-2448-1550},
B.~Adeva$^{49}$\lhcborcid{0000-0001-9756-3712},
M.~Adinolfi$^{57}$\lhcborcid{0000-0002-1326-1264},
P.~Adlarson$^{87,44}$\lhcborcid{0000-0001-6280-3851}
% ...

{\footnotesize \it

% ...
$^{39}$Nikhef National Institute for Subatomic Physics, Amsterdam, Netherlands\\
% ...
$^{44}$National Center for Nuclear Research (NCBJ), Warsaw, Poland\\
% ...
$^{49}$Instituto Galego de F{\'\i}sica de Altas Enerx{\'\i}as (IGFAE), Universidade de Santiago de Compostela, Santiago de Compostela, Spain\\
% ...
$^{53}$Physik-Institut, Universit{\"a}t Z{\"u}rich, Z{\"u}rich, Switzerland\\
% ...
$^{57}$H.H. Wills Physics Laboratory, University of Bristol, Bristol, United Kingdom\\
% ...
$^{59}$Department of Physics, University of Warwick, Coventry, United Kingdom\\
% ...
$^{61}$School of Physics and Astronomy, University of Edinburgh, Edinburgh, United Kingdom\\
$^{62}$School of Physics and Astronomy, University of Glasgow, Glasgow, United Kingdom\\
$^{63}$Oliver Lodge Laboratory, University of Liverpool, Liverpool, United Kingdom\\
% ...
$^{71}$Syracuse University, Syracuse, NY, United States\\
% ...
$^{87}$Department of Physics and Astronomy, Uppsala University, Uppsala, Sweden, associated to $^{62}$\\
% ...
}
\end{flushleft}

\end{document}